\documentclass[12pt]{article}
\usepackage{amsmath}
\usepackage{amsfonts}
\usepackage{amssymb}

\numberwithin{equation}{section}

\newcommand{\beq}{\begin{equation}}
\newcommand{\eeq}{\end{equation}}
\def\rf#1{(\ref{eq:#1})}
\def\Tr{\mathop{\rm Tr}}                  
\makeatletter
\newcommand{\rd}{\@ifnextchar^{\DIfF}{\DIfF^{}}}
\def\DIfF^#1{%
   \mathop{\mathrm{\mathstrut d}}%
   \nolimits^{#1}\gobblespace}
\def\gobblespace{\futurelet\diffarg\opspace}
\def\opspace{%
   \let\DiffSpace\!%
   \ifx\diffarg(%
   \let\DiffSpace\relax
   \else
   \ifx\diffarg[%
   \let\DiffSpace\relax
   \else
   \ifx\diffarg\{%
   \let\DiffSpace\relax
   \fi\fi\fi\DiffSpace}
\providecommand*{\dder}[3][]{%
\frac{\rd^{#1}#2}{\rd #3^{#1}}}
\providecommand*{\pder}[3][]{%
\frac{\partial^{#1}#2}{\partial #3^{#1}}}
\providecommand*{\iu}%
{\ensuremath{\mathrm{i}\,}}
\def\a{\alpha}
\def\b{\beta}

\def\d{\delta}
\def\g{\gamma}
\newcommand{\h}{\frac{1}{2}}

\begin{document}

\noindent
{\Large {\bf Solutions of the Painlev\'e VI Equation from Reduction of 
Integrable Hierarchy in a Grassmannian Approach}}
\vskip 9 mm
\begin{center}
\begin{minipage}[t]{70mm}
{\bf Henrik Aratyn}\\
\\
Department of Physics,\\
University of Illinois at Chicago,\\
845 W. Taylor St.,\\
Chicago, IL 60607-7059\\
e-mail: aratyn@uic.edu\\
\end{minipage}
\begin{minipage}[t]{70mm}
{\bf Johan van de Leur}\\
\\
Mathematical Institute,\\
University of Utrecht,\\
P.O. Box 80010, 3508 TA Utrecht,\\
The Netherlands\\
e-mail: J.W.vandeLeur@uu.nl
\end{minipage}
\end{center}

\begin{abstract}
We present an explicit method to perform similarity reduction 
of a Riemann-Hilbert factorization problem for a homogeneous 
$\widehat{GL} (N, {\mathbb C})$ loop group
and use our results to find solutions to 
the Painlev\'e VI equation for $N=3$.
The tau function of the reduced hierarchy 
is shown to satisfy the $\sigma$-form of 
the Painlev\'e VI equation.
A class of tau functions of the reduced integrable hierarchy 
is constructed by means of a Grassmannian formulation.
These solutions provide rational solutions of 
the Painlev\'e VI equation.
\end{abstract}

\section{Introduction}

One of main challenges in the study of integrable models
is derivation of a tau function providing solutions of 
the nonlinear partial differential hierarchy equations.
In this paper, we shall give a systematic and explicit construction 
of a self-similarity reduction of an infinite-dimensional integrable 
$\widehat{GL} (N, {\mathbb C})$  hierarchy \cite{Aratyn:2001cj,Aratyn:2004nz},
which in the case of $N=3$ describes a three-component
Kadomtsev-Petviashvili (KP) hierarchy \cite{KL}.
The   self-similarity reduction yields the $\sigma$-form of 
the Painlev\'e VI equation for which we explicitly construct a 
class of tau function solutions. 
We adopt Grassmannian approach to derive tau functions 
in terms of determinants obtained from 
expectation values of certain Fermi operators constructed 
using boson-fermion correspondence \cite{Aratyn:2006vc}. 
In particular,  we obtain a description of  
rational solutions of the Painlev\'e VI equation in terms of 
Schur polynomials.

In the context of soliton partial differential equations
a self-similarity reduction is closely related 
to the scaling behavior of their solutions.
To illustrate this relation let us examine an
example of the modified Korteweg-de Vries
(mKdV) equation  :
\[
u_t-6 u^2 u_x+u_{xxx}=0
\]
which has a self-similarity reduction \cite{Flaschka:1980wj} :
\beq
u (x,t)= (3 t)^{-1/3} f(z), \quad z= x (3 t)^{-1/3} \,.
\label{eq:pain2}
\eeq
Substitution of such function $u (x,t)$ into the mKdV equation
yields the following ordinary differential equation
\[
f^{\prime\prime}=2 f^3 +z f -\alpha, \quad \alpha={\rm constant}\, ,
\]
where $\;{}^{\prime}$ denotes derivative with respect to the appropriate
argument.
We recognize in the above relation the second Painlev\'e equation. 
It is of interest to point out that the single similarity variable
$z$ combines the two variables $x$ and $t$ in such a way 
as to ensure a scaling property :
\beq
u \left(\lambda x , \lambda^{3} t \right)= \lambda^{-1} u (x,t)\,, 
\label{eq:pain3}
\eeq
which characterizes function $u (x,t)$ in the self-similarity limit.
Alternatively one can reformulate the above scaling property 
as a linear condition $\left( x \partial_x +3 t \partial_t \right)
 u (x,t) = - u (x,t)$ being a special case of the so-called $L_{-1}$
 Virasoro constraint.

A common feature of soliton equations is that they are members
of integrable hierarchies, each hierarchy forms an infinite
sequence of evolution equations, labeled by their order.
In the example considered above the mKdV equation is the first 
member of the mKdV hierarchy while  the second Painlev\'e equation 
is the first member of the the second Painlev\'e hierarchy. 

The Riemann-Hilbert factorization problem 
serves the purpose of generating all 
evolution equations of the underlying hierarchy.
For the purpose of this paper we will use
the $\widehat{GL} (N, {\mathbb C})$  Riemann-Hilbert factorization 
problem of the form :
\begin{equation}
\exp \left( {\sum_{j=1}^{N} \sum_{n=1}^{\infty} 
z^n E_{jj}u^{(j)}_n}\right) \, g  (z) \,
= \Theta^{-1} ({ u},z)  \,  \Pi ({ u},z) \, . 
\label{eq:rh-def-gen}
\end{equation}
A self-similarity reduction (to which we will also refer
as a Painlev\'e reduction) is implemented by 
restricting $g:S^1 \to GL(N)$ to be of the special form :
\begin{equation}
g(z) = z^{ \underline{  \nu}}  \,g_0 \,z^{ -\underline{  \mu}} \, ,
\label{eq:gz-form}
\end{equation}
where $z^{ \underline{  \nu}}=
{\rm diag} ( z^{\nu_{1}}, {\ldots} , z^{\nu_{N}})$
and $z^{ -\underline{  \mu}}=
{\rm diag} ( z^{-\mu_{1}}, {\ldots} , z^{-\mu_{N}})$.
The reduction imposed by condition \rf{gz-form}
yields an integrable model parametrized by a set of conformal 
(scaling) dimensions $\mu_i, \nu_i\,,i=1,{\ldots} ,N$. 
The notation used in eq. \rf{rh-def-gen} and definitions 
of the dressing matrices $ \Theta({ u},z)$ and $\Pi({ u},z)$ will 
be explained in the next section.
Here we only note that $E_{ij}$ is a unit matrix 
with matrix elements given by 
$(E_{ij})_{kl} =\delta_{ik}\delta_{jl}$
and that the homogeneous gradation defined
by the gradation operator 
\[
\rd = z \dder{}{ z}
\]
is assumed. 

As seen below equation \rf{pain3} generators of the conformal 
symmetry naturally enter discussion of the self-similarity 
reduction process.
As indicated by \rf{pain3} an alternative mechanism to arrive at 
result of equation \rf{pain2} involves restricting 
$u (x,t)$ to be in the space of stationary solutions of an
appropriate  Virasoro symmetry generator 
\cite{Flaschka:1980wj,mazzoccomo} of the original hierarchy.
In a related method, the same result is obtained by imposing 
constraints on the Lax and Orlov-Schulman operators entering string equation
of the KdV hierarchy \cite{takasaki},
which when reformulated in terms of the wave-function can again
be expressed as a linear constraint involving action
of the Virasoro symmetry generator.
This shows equivalence of introducing a self-similarity
limit by either imposing  a certain scaling behavior
or a certain Virasoro condition.
Below, in equation \rf{vira-cond}, we will show
that the appropriate Virasoro condition ${\cal L}_{-1} \Psi  =
\underline{  \nu}  \Psi$ given in terms of the 
Virasoro operator ${\cal L}_{-1}$, characterizes 
 in an alternative way the reduction carried out in this paper.

The idea of imposing a well-defined scaling behavior on
dressing matrices to perform the self-similarity reduction of
the $\widehat{GL} (3, {\mathbb C})$ integrable hierarchy was 
already used recently by Kakei and Kikuchi \cite{kakei} who
established connection to $3 \times 3$ Fuchs--Garnier 
system \cite{Harnad,M2002} (see also \cite{joshi}) and 
subsequently to $2 \times 2 $ Schlesinger system of isomonodromy deformation 
flows known \cite{JM,Okamoto} to be associated with the Painlev\'e 
VI equation. 
In a related development, a one-dimensional reduction
to the generic Painlev\'e VI equation of the three-wave 
resonant system was constructed in \cite{conte}.
Here we provide a direct and explicit construction relating the 
tau function of the reduced  $\widehat{GL} (3, {\mathbb C})$
hierarchy to the solution of the $\sigma$-form of the 
Painlev\'e VI equation (see eq. \rf{jmo} below). 

To set the work in context let us point out that the 
construction in this paper generalizes a simpler case of the 
three-dimensional Frobenius manifold of the two-dimensional conformal field theory.
In that special case it was shown in \cite{Aratyn:2004nz}
that the above construction with a much simpler choice of 
$\underline{ \mu}={\rm diag} (-\mu,0,\mu)$ and 
$\underline{ \nu}=0$ leads to the Painlev\'e VI  equation  :
\begin{equation}
\begin{split}
 \dder[2]{y}{t} &=\frac{1}{2}\left( 
\frac{1}{y} + \frac{1}{y-1} + \frac{1}{y-t} 
\right)\left(\dder{y}{t}\right)^2
-\left(\frac{1}{t} + \frac{1}{t-1} + \frac{1}{y-t} 
 \right)\dder{y}{t}\\
&+ \frac{y(y-1)(y-t)}{t^2(t-1)^2}
\left\{ \alpha + \beta \frac{t}{y^2}
+ \gamma \frac{t-1}{(y-1)^2}
+ \delta \frac{t(t-1)}{(y-t)^2} \right\},
\end{split}
\label{eq:P6}
\end{equation}
with the Painlev\'e parameters $(\a,\b,\g,\d)$
fully determined by a single conformal
scaling parameter $\mu$ through :
\begin{equation}
\alpha= \frac{(1\mp \mu)^2}{2},\;\;\; \beta = - \frac{\mu^2}{2},\;\;\;
\gamma= \frac{\mu^2}{2},\;\;\; \delta = \frac{1-\mu^2}{2},
\label{eq:reduca}
\end{equation}
Note that, in an equivalent parametrization of
\cite{Dubrovin:1998fe,Aratyn:2004nz},
the Painlev\'e parameters $(\a,\b,\g,\d)$ take the following form :
\begin{equation}
\alpha= \frac{(1 \pm 2\mu)^2}{2},\;\;\; \beta = 0,\;\;\;
\gamma= 0,\;\;\; \delta = \frac{1}{2} \, .
\label{eq:reducb}
\end{equation}
The above restriction on the allowed values of  the Painlev\'e 
parameters $(\a,\b,\g,\d)$ no longer holds in a more general 
setting defined by \rf{gz-form} and, as shown below, we find 
relations of the type
\begin{alignat}{2} 
\b&=-\frac{(\nu_2-\mu_3)^2}{2},&\qquad 
\g=\frac{(\mu_1-\mu_2)^2}{2} \nonumber  \\
\a&=\frac{(\nu_3-\mu_3)^2}{2}
,&\qquad
\d= \frac{1}{2} \left(1 -\left( 1+\nu_1-\mu_3\right)^2\right)\, .
\nonumber
\end{alignat}
For $g(z) $ as in \rf{gz-form} and $N=3$ we will show that 
certain dressing matrix can be transformed by a matrix similarity 
transformation
in such a way that it is determined by six variables
$\omega_i, {\bar \omega}_i, \, i=1,2,3$, which 
only depend on one single variable 
\[
t=\frac{u_1^{(2)}-u_1^{(1)}}{u_1^{(3)}-u_1^{(1)}} \, .
\]
These six variables satisfy generalizations 
the familiar Euler top equations:
\begin{alignat}{2}
\pder{}{t} \omega_3 &= \frac{\bar{\omega}_1\omega_2}{1-t}
 ,&\quad \pder{}{t} \bar{\omega}_3 &= \frac{\omega_1 \bar{\omega}_2}{1-t}
\nonumber\\
\pder{}{t} \omega_2 &=  \frac{\omega_1\omega_3}{t(t-1)} - \frac{ \omega_2}{t}
(\nu_3-\nu_1), &\quad 
\pder{}{t} \bar{\omega}_2 &=  \frac{\bar{\omega}_1\bar{\omega}_3}{t(t-1)} 
+ \frac{ \bar{\omega}_2}{t} (\nu_3-\nu_1) \label{eq:omeqs}\\
\pder{}{t} \omega_1  &= \frac{\omega_2\bar{\omega}_3}{t}
+ \frac{\omega_1}{t(t-1)} (\nu_3-\nu_2), &\quad 
\pder{}{t} \bar{\omega}_1  &= \frac{\bar{\omega}_2\omega_3}{t}
- \frac{\bar{\omega}_1}{t(t-1)} (\nu_3-\nu_2) \nonumber
\end{alignat}
and are constrained by a quadratic 
algebraic relation $\sum_i \omega_i {\bar \omega}_i =-R^2$ with
a constant $R$ defined in \rf{Roms} as well as 
a cubic algebraic relation given in relation \rf{omsboms}.

Equivalence is established between on the one hand the above system of 
equations \rf{omeqs} and the algebraic constraints for arbitrary
values of scaling parameters $\mu_i,\nu_i,\,i=1,2,3$
and on the other hand the Painlev\'e VI equation \rf{P6}
with the complete set of the Painlev\'e parameters $(\a,\b,\g,\d)$.
Explicit construction shows that the system consisting of relations 
\[\omega_2 {\bar \omega}_2= f^\prime, \;\;\; 
\omega_1 {\bar \omega}_1=-f+(t-1)  f^\prime,
\]
between $\omega$'s and the tau function $\tau$ 
represented by $ f(t) = t(t-1) d \log \tau /dt$ 
and the algebraic constraints for $\omega$'s is equivalent to 
the so-called $\sigma$-form of the Painlev\'e VI
equation \cite{JM}:
\begin{equation}
\dder{\sigma}{t} \left(t (t-1) \dder[2]{\sigma}{t}\right)^2+
\left( \dder{\sigma}{t}  \left[2 \sigma
-(2t-1) \dder{\sigma}{t}\right] +v_1v_2v_3v_4\right)^2
=\prod_{k=1}^4 \left( \dder{\sigma}{t}+v_k^2\right)\, , 
\label{eq:jmo}
\end{equation}
where, for  certain constants $a,b$,  
\[
\sigma=t(t-1)\frac{d \log \tau}{dt}-a t - b\, .
\]
Note, that a similar relation between the tau function underlying the Painlev\'e VI
system and $\sigma$ appeared in \cite{Okamoto} with different
values of $a$ and $b$, which implies that our KP derived tau function
differs from the tau function used in \cite{Okamoto}.

The $\sigma$-form of the Painlev\'e VI equation exhibits 
a $D_4$ root system symmetry in the parameters 
$v_1 , {\ldots} , v_4$, which are related to
the general Painlev\'e parameters $(\a,\b,\g,\d)$ through
\begin{equation}
v_1+v_2=\sqrt{-2\beta},\;\; v_1-v_2=\sqrt{2\gamma},\;\;
v_3+v_4+1=\sqrt{1-2\delta},\;\; v_3-v_4=\sqrt{2\alpha}\, .
\label{eq:via}
\end{equation}
Below, we will find the realization of the above parameters in terms
of the scaling dimensions $\mu_i, \nu_i\, ,i=1,{\ldots} ,3$.

In the $3$-component Clifford algebra setting the tau-function
is explicitly realized as 
\[
\tau( \nu_1, \nu_2,\nu_3;{\bf u})=
\langle 0|Q_3^{-\nu_3}Q_2^{-\nu_2}Q_1^{-\nu_1}e^{\sum_{i,k}\alpha^{(i)}u_k^{(i)}}
G|0\rangle
\, ,
\]
where, as explained below, the element
$
Q_3^{-\nu_3}Q_2^{-\nu_2}Q_1^{-\nu_1}$ is related to 
$z^{\underline \nu}$, $e^{\sum_{i,k}\alpha_k^{(i)}u_k^{(i)}}$ to
$e^{\sum_{i,k}z^iE_{kk}u_k^{(i)}}$ and $G|0\rangle$ corresponds to
$g_0 z^{-\underline\mu}$ (cf. (\ref{eq:gz-form})) 
acting on the vacuum 
\[
|0\rangle=e_3\wedge e_2\wedge e_1\wedge ze_3\wedge ze_2\wedge ze_1\wedge z^2e_3
\wedge z^2e_2\wedge z^2e_1\wedge z^3e_3\wedge z^3e_2\wedge z^3e_1\wedge
z^4e_3\wedge \cdots \, , 
\]
where $\{ e_i\}_{1\le i\le 3}$ is a basis of $\mathbb{C}^3$.

The paper is organized as follows.
The main subject of Section 2 is a ${\widehat{GL} (N, {\mathbb C})}$ 
hierarchy in the self-similarity limit characterized by the scaling parameters
$\underline{ \mu}$ and $\underline{ \nu}$. 
In the end of this section the ${ \sigma}$-form of the Painlev\'e 
VI equation is obtained for the underlying tau function 
and the parameters of the Painlev\'e VI equation are related to 
the scaling parameters $\underline{ \mu}$ and $\underline{ \nu}$. 
Section 3 develops an explicit Grassmannian realization
of Riemann­-Hilbert factorization problem leading to 
the tau function given as an expectation value
within the 3-component  Clifford algebra setting. 
Some background material about semi-infinite wedge space 
and Clifford algebra is also given here.
Section 4 derives conditions satisfied by the tau function constructed in the
previous section.
In the next section, Section 5, the tau function is cast in the
determinant form. The technical details of this Section are 
relocated to Appendix A.
The wave matrix is calculated in Section 7.
In Section 6 we describe an explicit example, which produces e.g. the solution
\begin{equation}
\label{y}
y\, = \,
{\frac {-{{\it D_2}}^{2}{t}^{5}+3\,{{\it D_2}}^{2}{t}^{4}+2\,{\it D_1}
\,{\it D_2}\,{t}^{3}+2\,{\it D_1}\,{\it D_2}\,{t}^{2}+
3\,{{\it D_1}}^{2}\,t-{{\it D_1}}^{2}}{{{\it D_1}}^{2}+
4\,{\it D_2}\,{\it D_1}\,t-
6\,{\it D_1}\,{\it D_2}\,{t}^{2}+
4\,{\it D_1}\,{\it D_2}\,{t}^{3}+{{\it D_2}}^{2}{t}^{4}\\
\mbox{}}}
\end{equation}
of the Painlev\'e VI equation
(\ref{eq:P6}) for arbitrary constants $D_1,D_2\in\mathbb{C}$ and the parameters 
\[
\alpha=\frac12,\quad \beta=-2,\quad \gamma=2,
\quad\mbox{ and }\quad\delta=-\frac32\, .
\]

\section{${\widehat{GL} (N, {\mathbb C})}$ Hierarchy in the Self-Similarity Limit}
Let the Riemann-Hilbert factorization problem for 
$\widehat{GL} (N, {\mathbb C})$
be defined as in equations \rf{rh-def-gen} and \rf{gz-form}
with all higher flows $u^{(j)}_n,\, n>1$ 
set to zero. For notational convenience we set
$u_j=u^{(j)}_1, \;j=1,{\ldots} ,N$ and use the multi-time 
notation with ${ u}= (u_1, {\ldots} , u_{N})$
to denote all $N$ $u_j$-flows.
For brevity we sometimes denote
$ \partial_j = {\partial}/ {\partial u_j}$ for
$j=1,{\ldots}, N$.
(Later on we will also use the notation $\bf u$ for all $u_i^{(k)}$, $i=1,2,3,\ldots$,
$k=1,2,3$, when we do not put all higher flows $u^{(k)}_i,\, i>1$ 
to zero.)

Equation \rf{rh-def-gen} simplifies to 
\begin{equation}
\exp \left( {\sum_{j=1}^{N} z E_{jj}u _j}\right) \, g  (z) \,
= \Theta^{-1} ({ u},z) \,  \Pi  ({ u},z), 
\label{eq:rh-def}
\end{equation}


The dressing matrices $ \Theta$, $\Pi$ in equation \rf{rh-def} 
have the following expansions into positive and negative modes 
with respect to the $\rd$ gradation operator :
\begin{eqnarray}
\Theta ({ u},z)&=& 1 + \theta^{(-1)} ({ u})z^{-1} +
\theta^{(-2)}  ({ u}) z^{-2}+{\ldots} 
\label{eq:tgrad-exp}\\
\Pi ({ u},z)&=& M ({ u})\left(1 + \pi^{(1)} ({ u}) z + \pi^{(2)} ({ u}) z^{2}+{\ldots}\right) \,\,.
\label{eq:pgrad-exp}
\end{eqnarray}

One derives 
from \rf{rh-def} the following expressions for the symmetry 
$u_j$-flows  :
\begin{eqnarray}
\frac{\partial}{\partial u_j} \Theta ({ u}, z) 
&=& - \left(\Theta z E_{jj} \Theta^{-1} 
\right)_{-} \Theta ({ u}, z) 
\label{eq:uthpos}\\
\frac{\partial}{\partial u_j} \Pi ({ u}, z) 
&=&  \left(\Theta z E_{jj} \Theta^{-1} 
\right)_{+} \Pi ({ u}, z) 
\label{eq:ummpos}
\end{eqnarray}
where $({\ldots} )_{\pm}$ denote the projections onto 
the negative and positive powers of $z$, respectively.

We now choose $g:S^1 \to GL(N)$ to be given by relation \rf{gz-form}
so that the Riemann-Hilbert factorization problem 
becomes
\begin{equation}
\begin{split}
&e^{  z \,\underline{ u}\,}  \; z^{\, \underline{  \nu}}\; g_0 \;z^{-\, \underline{  \mu}}\,
= \Theta^{-1} ({ u},z) \,  \Pi  ({ u},z) \\
&\underline{  \nu}=  \sum_{j=1}^N \nu_j E_{jj}, \qquad
\underline{  \mu}=\sum_{j=1}^N \mu_j E_{jj} , \qquad
\underline{ u}\, =\sum_{j=1}^{N} u_j E_{jj}
\end{split}
\label{eq:rh-numu}
\end{equation}
where $g_0$ is a grade zero element.
The Riemann-Hilbert factorization problem \rf{rh-numu}
leads to the $u_j$-flows of the same functional form
as in \rf{uthpos} and  \rf{ummpos} as they are not affected of the functional
form of $g(z)$.

Applying the grading operator $\rd$ on both 
sides of eq. \rf{rh-numu} one finds for the negative and strictly
positive grades:
\begin{equation}
\begin{split}
\rd \Theta&=  - \left(\Theta z \,\underline{ u}\, \Theta^{-1} 
\right)_{-} \Theta  - \left(\Theta \, \underline{  \nu} \Theta^{-1} 
\right)_{-} \Theta =- \left(\Theta z \,\underline{ u}\, \Theta^{-1} 
\right)_{-} \Theta  - \Theta \, \underline{  \nu} 
+  \underline{  \nu} \Theta \\
\rd \Pi &=   \left(\Theta z \,\underline{ u}\, \Theta^{-1} 
\right)_{>0} \Pi  - \left(\Pi \, \underline{  \mu} \Pi^{-1} 
\right)_{>0} \Pi 
\end{split}
\label{eq:dtheta}
\end{equation}
For the grade zero one obtains from equation \rf{rh-numu}
a consistency condition:
\begin{equation}
\left(\Theta z \,\underline{ u}\, \Theta^{-1} 
\right)_{0}
+ \left(\Theta \, \underline{  \nu} \Theta^{-1} 
\right)_{0}
-\left(\Pi \, \underline{  \mu} \Pi^{-1} 
\right)_{0} 
= \left(\Theta z \,\underline{ u}\, \Theta^{-1} 
\right)_{0}+\, \underline{  \nu}-  M \, \underline{  \mu} M^{-1} =0
\label{eq:dcondition}
\end{equation}
It is convenient to define the unity
$\widehat{I}$ and the Euler $\widehat{E}$ vector fields as:
\begin{equation}
\widehat{I} = \sum_{j=1}^{N}   \frac{\partial}{ \partial u_j}, \qquad 
 \widehat{E} = \sum_{j=1}^{N}  u_j
 \frac{\partial}{ \partial u_j}
\label{eq:euler}
\end{equation}
Note that from \rf{dtheta} and \rf{uthpos} one finds
\begin{equation}
\left( z \pder{}{z} - {\widehat E} \right)  \Theta ({ u}, z) = 
-\left(\Theta \underline{ \nu} \Theta^{-1} \right)_{-}\Theta
= \left\lbrack \underline{ \nu} , \Theta \right\rbrack , \qquad
{\widehat I} \left( \Theta \right) =0\,.
\label{eq:confrob}
\end{equation}
Relation \rf{confrob} implies the following scaling law for
$\Theta$:
\begin{equation}
 \Theta ({ u}, \lambda z) = \lambda^{\, \underline{  \nu}}\; \Theta (
 \lambda \, { u}, 
 z)\; \lambda^{\,-\underline{  \nu}} \, .
\label{eq:th-scale}
\end{equation}

{}From eq. \rf{uthpos} it follows quite generally 
for the diagonal elements of $\theta^{(-1)}$ :
\[
\begin{split}
\partial_j (\theta^{(-1)})_{ii} &= - \left(\Theta z E_{jj} \Theta^{-1} 
\right)_{-1\, ii} 
=   \left(\theta^{(-1)}\, E_{jj} \theta^{(-1)}\right)_{ii}
\\
&=  \b_{ij}\,\b_{ji}\;\;\;\; i\ne j=1,{\ldots} ,N
\end{split}
\]
where we introduced the ``rotation coefficients'' $\b_{ij}$ with $i\ne j$ as
the off-diagonal elements of the $\theta^{(-1)}$ matrix :
\begin{equation}
\b_{ij}({ u})\, =\, (\theta^{(-1)})_{ij} ({ u}), \;\;\;\; i\ne
j=1,{\ldots} ,N\, .
\label{eq:them1}
\end{equation}
Thus,
\[
\partial_j (\theta^{(-1)})_{ii}-\partial_i (\theta^{(-1)})_{jj}=0
\]
and accordingly we can express the diagonal elements 
of the $\theta^{(-1)}$ matrix as a derivative of a logarithm of a
tau-function :
\begin{equation}
(\theta^{(-1)})_{ii} = - \partial_i  \log \tau \,.
\label{eq:thm1t}
\end{equation}
It follows that
\begin{equation}
\partial_j \partial_i  \log \tau  = - \b_{ij}\,\b_{ji}\;\;\;\; i\ne j=1,{\ldots} ,N
\label{eq:taubetas}
\end{equation}
For $i=j$ in the above equation we get from eq. \rf{uthpos} :
\[
\begin{split}
\partial_i (\theta^{(-1)})_{ii} &= - \left(\Theta z E_{ii} \Theta^{-1} 
\right)_{-1\, ii} 
=   \left(\theta^{(-1)}\, E_{ii} \theta^{(-1)}\right)_{ii}
- \left(\theta^{(-1)}\right)^2_{ii}
\\
&=  (\theta^{(-1)}_{ii})^2- \left(\theta^{(-1)}\right)^2_{ii}
=- \sum_{k \ne i} \b_{ik}\,\b_{ki}
\end{split}
\]
To summarize, we have found that
\begin{equation}
\partial_i \partial_j  \log \tau = 
\left\{ \begin{matrix}
-\b_{ij} \b_{ji} & i \ne j\\
 \sum_{k \ne i} \b_{ik}\,\b_{ki} & i=j\, .
\end{matrix} \right.
\label{eq:pijtau}
\end{equation}
{}From eq. \rf{th-scale} we derive
\begin{equation}
\theta^{(-1)} ({ u}) = \lambda^{1+ \underline{  \nu}}\; \theta^{(-1)} 
(\lambda \, { u})\; \lambda^{\,-\underline{  \nu}} 
\label{eq:theta1scale}
\end{equation}
or for the matrix-elements of $\theta^{(-1)} ({ u})$ :
\begin{equation} 
\widehat{ E}  \left(\theta^{(-1)}_{ij}\right)= -(1+\nu_i-\nu_j)
\theta^{(-1)}_{ij} \quad
{\rm or} \quad \widehat{ E} \left( \b_{ij}\right)= -(1+\nu_i-\nu_j) \b_{ij}
\label{eq:t1e}
\end{equation}
The above results also follow from eq. \rf{confrob}. 
In particular, for the diagonal elements of the dressing
matrix one gets :
\[
\widehat{ E} \left( \theta^{(-1)}_{ii}\right) =- \theta^{(-1)}_{ii}\quad
\to\quad \widehat{ E} \left( \partial_i \log \tau \right) = -  \partial_i \log \tau 
\quad
\to \quad \partial_i \widehat{ E} \left( \log \tau \right)=0
\]
which amounts to
\[
 \widehat{ E} \left(\log \tau\right) = {\rm const} \, .
\]
Using eqs. \rf{taubetas} and \rf{pijtau}
we obtain
\[
\begin{split}
\widehat{ E} \left( \partial_i \log \tau \right) &=
\sum_{j=1}^{N} u_{j} \partial_{j} \partial_{i} \log \tau
= -\sum_{j \ne i}^{N} u_{j} \b_{ij}\,\b_{ji}
+ u_{i}\sum_{k \ne i} \b_{ik}\,\b_{ki}  \\
&= -\sum_{j=1}^{N} \left(u_{j}-u_{i} \right) \b_{ij}\,\b_{ji}
\end{split}
\]
It follows from 
$0= \partial_i \widehat{ E} \left( \log \tau \right)
=  \partial_i \log \tau   +\widehat{ E} \left( \partial_i \log \tau \right)$
that
\[
\partial_i \log \tau= - \theta^{(-1)}_{ii}=\sum_{j=1}^{N} \left(u_{j}-u_{i}
\right) \b_{ij}\,\b_{ji} \, .
\]
Thus
\[
\widehat{ I} \left( \log \tau \right)= \sum_{i,j=1}^{N} \left(u_{j}-u_{i} \right) \b_{ij}\,\b_{ji}
=0
\]
and
\begin{equation}
\begin{split}
\widehat{ E} \left( \log \tau \right)&=\sum_{i=1}^{N} u_{i}\partial_i \log \tau=
\sum_{i,j=1}^{N} u_{i} \left(u_{j}-u_{i} \right) \b_{ij}\,\b_{ji}\\
&=-\b_{12}\b_{21} \left(u_{1}-u_{2}\right)^{2} -\b_{13}\b_{31} \left(u_{1}-u_{3}\right)^{2}
  -\b_{32}\b_{23} \left(u_{3}-u_{2}\right)^{2}\\
&= V_{12} V_{21}+V_{13} V_{31}+ V_{32}V_{23}= \h \Tr \left(V^{2}\right)
\quad {\rm for} \;\;\;N=3\,,
  \end{split}
\label{eq:tracev2}  
\end{equation}
where we have introduced
\begin{equation}
V \equiv \lbrack \theta^{(-1)} \, , \, \,\underline{ u}\, \rbrack =\left(\Theta z \,\underline{ u}\, \Theta^{-1} 
\right)_{0} \,.
\label{eq:v-def}
\end{equation}
Matrix $V$ reads in components :
\begin{equation}
V_{ij} = (u_j-u_i) \theta^{(-1)}_{ij}=(u_j-u_i) \b_{ij} , \quad
i,j=1, {\ldots} ,N \,.
\label{eq:v-comps}
\end{equation}
Thus we conclude that $\Tr \left(V^{2}\right)$ is a constant.
Below we will find that $\Tr \left(V^{2}\right)=\sum_i \mu_i^2 - \sum_i \nu_i^2$.

Applying the scaling law \rf{th-scale} to expressions 
\rf{them1}, \rf{thm1t} we obtain scaling rules 
\begin{equation}
\tau (\lambda \, { u}) =  \tau ({ u}),
\qquad
\b_{i\,j} (\lambda \, { u}) = \lambda^{-(1+\nu_i-\nu_j)} \b_{i\,j} ({ u}),
\label{eq:taub}
\end{equation}
which are consistent with relations \rf{t1e}.

The similar scaling law for $\Pi$ can be obtained from
second of eqs. \rf{dtheta}. First using relation \rf{dcondition}
we obtain 
\[
\rd\, \Pi =   \left(\Theta z \,\underline{ u}\, \Theta^{-1} 
\right)_{+} \Pi  + \left(\Theta \, \underline{  \nu} \Theta^{-1} 
\right)_{0} - \left(\Pi \, \underline{  \mu} \Pi^{-1} 
\right)_{+} \Pi \, ,
\]
which leads to 
\begin{equation}
\left( \rd  - \widehat{E} \right)  \Pi ({ u}, z)  
= \underline{ \nu}  \Pi ({ u}, z) -  \Pi ({ u}, z) \underline{ \mu}
\label{eq:confpi}
\end{equation}
consistent with the following scaling law 
\begin{equation}
\Pi ({ u}, \lambda z) = \lambda^{\, \underline{  \nu}} \;\Pi (\lambda \, { u}, 
 z)\; \lambda^{\,-\underline{  \mu}} \, .
\label{eq:pi-scale}
\end{equation}

{}From eqs. \rf{confpi} and \rf{pi-scale} it 
clearly follows that
\begin{equation}
\widehat{E}( M ) ({ u})= M({ u}) \underline{ \mu} -\underline{ \nu} M({ u})
, \qquad
M({ u})=  \lambda^{\, \underline{  \nu}} \;M (\lambda \, { u})\; \lambda^{\,-\underline{  \mu}} \, .
\label{eq:M-trans}
\end{equation}

Define a wave matrix $\Psi$ as :
\begin{equation}
\Psi ({ u}, z) = \Theta ({ u}, z)\, 
e^{z\,\underline{ u}\,}  z^{\, \underline{  \nu}} \, .
\label{eq:gms}
\end{equation}
Then
\begin{equation}
\rd \,\Psi= \left( \left(\Theta (z\,\underline{ u}\,) \Theta^{-1} 
\right)_{+}+ \, \underline{  \nu} \right) \Psi
\label{eq:dpsi}
\end{equation}
and
\[ \frac{\partial }{\partial u_i} \Psi (z,t) =  
\left(\Theta z E_{ii} \Theta^{-1}\right)_{+}
\Psi (z,t) 
\]
Thus
\begin{equation}
  \left( \rd  - \widehat{E} \right)  \Psi ({ u}, z) =
\underline{  \nu}  \Psi ({ u}, z)      \, ,
\label{eq:vira-cond}
\end{equation}
which ensures the scaling behavior $\Psi ({ u}, \lambda z) 
= \lambda^{\, \underline{  \nu}} \; \Psi (\lambda \, { u},  z) $.

The operator on the right hand side was identified with the 
Virasoro operator ${\cal L}_{-1}$ in \cite{Aratyn:2001jv},
where the corresponding left hand side of the Virasoro
condition vanished.

\subsection{Reduction. ${\widehat{GL} (3, {\mathbb C})}$ Hierarchy.}
{}From now on we set $N=3$.  

The scaling law for the matrix elements of $V$ reads 
\[
V_{ij} ({ u}) = \lambda^{\nu_i- \nu_j}\;  V_{ij} (\lambda \, { u})
\]
In terms of  the matrix
$V $  condition \rf{dcondition} 
becomes :
\begin{equation}
V +\, \underline{  \nu}-  M \, \underline{  \mu} M^{-1} =0
\label{eq:dcondred}
\end{equation}
or
\[ V M = M \underline{ \mu} - \underline{ \nu} M = \widehat{E}( M) \, .
\]
Eq. \rf{dpsi} can be written as : 
\begin{equation}
\rd\, \Psi= \left( z \,\underline{ u}\, +V + \, \underline{  \nu} \right) \Psi
\label{eq:dpsired}
\end{equation}
Furthermore
\begin{equation}
\frac{\partial \Psi}{\partial u_j} =(z E_{jj}+V_j) \Psi \, .
\label{eq:cdjpsi}
\end{equation}
and 
\begin{equation} 
\frac{\partial }{\partial u_j} M =V_j M \, .
\label{eq:cdjm}
\end{equation}
where we introduced a matrix
\begin{equation}
V_j \equiv \lbrack \theta^{(-1)} \, , \, E_{jj}\rbrack , \;\;\;
(V_j)_{kl}  = \left( \d_{lj} -  \d_{kj}\right)  \theta^{(-1)}_{kl} \, ,
\label{eq:defvi}
\end{equation}
which scales as in \rf{theta1scale}.

{}From compatibility of the above equations it follows that
\begin{equation}
\begin{split}
\partial_j V &= \lbrack V_j  \, , \, V + \, \underline{  \nu}\rbrack  \\
\partial_j V_i &= \partial_i V_j +\lbrack V_j  \, , \, V_i\rbrack  \\
 \lbrack V  \, , \, E_{jj}\rbrack &= \lbrack V_j  \, , \, \,\underline{ u}\,\rbrack 
\end{split}
\label{eq:pjv}
\end{equation}
Also, since diagonal elements of 
$V = \lbrack \theta^{(-1)} \, , \, \,\underline{ u}\, \rbrack$
are zero, it follows from 
\rf{dcondred} that
\begin{equation}
\nu_i = \left(M \, \underline{  \mu} M^{-1} \right)_{ii}
\, , \quad i=1, {\ldots} ,N
\label{eq:nudiag}
\end{equation}
Taking trace on both side we obtain a trace condition :
\begin{equation}
\label{trace-cond}
\nu_1 +\nu_2+ \nu_3 = \mu_1 +\mu_2+ \mu_3
\end{equation}
The special case of $\underline{ \mu}=-\mu,0,\mu$ 
and $\underline{ \nu}=0$ was already considered in \cite{Aratyn:2004nz}.

It holds that
\begin{equation}
\widehat{I} (V)=0, \quad \widehat{E} (V)=  \left\lbrack V \, ,\,  \underline{ \nu}  \right\rbrack
, \quad \widehat{E} \left( M \right)=V  M \, .
\end{equation}
The identity $\widehat{I} (V)=0$, shows that $V$ is a function of two
variables. Those can be identified with $t$ and $h$ given by:
\begin{equation}
\label{th}
t= \frac{u_2-u_1}{u_3-u_1}\,, \qquad
h=u_2-u_1 \, .
\end{equation}
and thus $V ({ u}) = V(t,h)$.
Making use of technical identities :
\[
\pder{t}{u_1}=\frac{1}{h} (t-1)t, \quad
\pder{t}{u_2}=\frac{1}{h} t, \quad
\pder{t}{u_3}=- \frac{1}{h} t^2 ,
\]
one easily derives
\[
\pder{}{u_1} = \frac{t(t-1)}{h} \pder{}{t}-\pder{}{h},\;\;\; \,
\pder{}{u_2}= \frac{t}{h}\pder{}{t}+\pder{}{h},\;\;\; \,
\pder{}{u_3}=-\frac{t^2}{h}\pder{}{t}\, ,
\]
{}from which
\[
\widehat{E}= h \pder{}{h}\,
\]
follows.

Now, define 
\begin{equation}
\bar{V}= e^{\underline{ \nu}  \log h}\, V (t,h) \, e^{- \underline{ \nu} \log h}
=\lbrack     {\bar \theta}^{(-1)} \, , \, \,\underline{ u}\, \rbrack \, ,
\label{eq:Vbar}
\end{equation}
where
\begin{equation}
{\bar \theta}^{(-1)}= e^{\underline{ \nu}  \log h}\, \theta^{(-1)} \, e^{- \underline{ \nu} \log h}
\label{eq:thetabar}
\end{equation}
has scaling dimension one.

$\bar{V}$ satisfies 
\[
\widehat{E} \left( \bar{V} \right) 
= h \pder{\bar{V}}{h} = 
\, \left\lbrack  \underline{ \nu} \, ,\,  \bar{V}  \right\rbrack+
e^{\underline{ \nu} \log h} \, \left\lbrack V \, ,\,  \underline{ \nu}  
\right\rbrack  e^{- \underline{ \nu} \log h}
= 0
\]
Because of $\widehat{I} (h)=0$ it also follows that 
$\widehat{I} (\bar{V})=0$ and, thus, the matrix
$\bar{V}$ is a function of  only one variable $t$:
$\bar{V}=\bar{V} (t)$.

Similarly define,
\begin{equation} 
\bar{M}= e^{\underline{ \nu}  \log h}\, M\, e^{-\underline{ \mu}  \log h}
\label{eq:Mbar}
\end{equation}
Then
\begin{equation}
\begin{split}
\bar{V} + \underline{ \nu} &= \bar{M}\underline{ \mu} \bar{M}^{-1} = e^{\underline{ \nu}  \log h}\, {M}\underline{ \mu} {M}^{-1} e^{-\underline{ \nu}  \log h}
\\
\widehat{E} \left( \bar{M} \right) &= \left( \bar{V} + \underline{ \nu}\right)  \bar{M}
-  \bar{M} \underline{ \mu} = 0
\end{split}
\label{eq:barmrels}
\end{equation}
Thus, the matrix $\bar{M}$ is also 
 a function of only one variable $t$ (since it follows from
\rf{cdjm} that $\widehat{I} (\bar{M})=0$) and  we set 
$\bar{M}=\bar{M} (t)$.

We will use the following parametrization of 
matrices from eq. \rf{barmrels}
\begin{equation}
\bar{V} + \underline{ \nu} = \bar{M}\underline{ \mu} \bar{M}^{-1}
= \begin{bmatrix} \nu_1 & \omega_3 & -\omega_2 \\
-\bar{\omega}_3 & \nu_2 & \omega_1 \\
\bar{\omega}_2 & -\bar{\omega}_1& \nu_1
\end{bmatrix}
\label{eq:omegas}
\end{equation}

{}From the first of equations \rf{pjv} we find
\[
\pder{}{t} \bar{V} = - \frac{h}{t^2} \lbrack \bar{V}_3 \,, \, 
\bar{V} + \underline{ \nu} \rbrack, \quad
\bar{V}_j =  \lbrack    e^{\underline{ \nu}  \log h}\, \theta^{(-1)} \, e^{- \underline{ \nu} \log h}
\, , \, E_{jj} \rbrack \, ,
\]
which leads to equations  \rf{omeqs} for
the matrix elements from \rf{omegas}.
The basic observation to make in connection
with  equations \rf{omeqs} is that the limit $\omega_i = \bar{\omega}_i$ 
is not consistent with these equations unless 
$\nu_3=\nu_2=\nu_1$. In such limit  $V$ is skew-symmetric
and $M$ is orthogonal.

Next, consider $\Tr (V^2)=\Tr ({\bar V}^2) =\sum_i \omega_i \bar{\omega}_i$. From equations
\rf{omeqs} it follows that
\[
\pder{}{t} \sum_i \omega_i \bar{\omega}_i = 0\, ,
\]
in agreement with relation \rf{tracev2}.
A direct calculation gives from eq. \rf{dcondred}
\[
\Tr \left((V +\, \underline{  \nu})^2\right) = \Tr \left(\underline{  \mu}^2\right) 
\]
or
\[
\Tr \left(V^2\right) = \Tr ({\bar V}^2) =\Tr \left(\underline{  \mu}^2\right) - \Tr \left(\underline{  \nu}^2\right) 
\]
which indeed is a constant.

In addition from
the trace relation
$ \Tr \left((V+\nu)^{2}\right)=\sum_{i}( \nu_{i}^{2} - 2 \omega_i {\bar \omega}_i)$
we have a condition:
\begin{equation}
\omega_1 {\bar \omega}_1+ \omega_2 {\bar \omega}_2+ \omega_3 {\bar \omega}_3
=- \h(\mu_1^2+\mu_2^2+\mu_3^2- \nu_1^2-\nu_2^2-\nu_3^2) = -R^2\, ,
\label{eq:Roms}
\end{equation}
where for convenience  we have introduced on the right hand side
a new constant $R$. Since the trace condition (\ref{trace-cond}) holds, one finds for 
$c\in\mathbb{C}$ that
\[
R^2=\h\left((\mu_1-c)^2+(\mu_2-c)^2+(\mu_3-c)^2-(\nu_1-c)^2-(\nu_2-c)^2-(\nu_3-c)^2\right)\, .
\]
\subsection{The tau function and the ${ \sigma}$-form of the Painlev\'e VI
equation}

We can use matrices $V$ and $V_{j}$ to find an alternative 
expressions for derivatives of the tau function as:
\begin{equation}
\partial_j \log \tau   = 
\h \Tr (\bar{V}_j \bar{V}) , \qquad \partial_i \partial_j \log \tau  
=-\h \Tr ( \bar{V}_i \bar{V}_j) ,\qquad j=1,2,3 \, ,
\label{eq:tau1}
\end{equation}  
These identities allow us to express the right hand side of \rf{pijtau}
by $\omega_i , \bar{\omega}_i$.
First note that from eq. \rf{v-comps} rewritten in a ``barred''
version it follows that
\begin{alignat}{2}  
\bar{\b}_{12} & =\frac{\omega_3}{h}, & \qquad \bar{\b}_{21} & 
=\frac{{\bar\omega}_3}{h}\, ,
\nonumber\\  
\bar{\b}_{13} & =-\frac{t\omega_2}{h}, & \qquad \bar{\b}_{31} & =-\frac{t{\bar\omega}_2}{h}\, ,
\nonumber\\  
\bar{\b}_{23} & =\frac{t\omega_1}{h(1-t)}, & \qquad \bar{\b}_{32} &
=\frac{t{\bar\omega}_1}{h(1-t)} \, .
\nonumber
\end{alignat}
{} Furthermore from
\[
\begin{split}
\frac{\partial^2 }{\partial u_1 \partial u_3} &= - \frac{t^2}{h^2} \left[ t (t-1) 
\pder[2]{}{t}+(2t-1) \pder{}{t}-h\pder{}{h}\pder{}{t} \right]\\
\frac{\partial^2 }{\partial u_1 \partial u_2} &=  \frac{1}{h^2} \left[ t^{2} (t-1) 
\pder[2]{}{t}+t^{2} \pder{}{t}+h(t^{2}-2t)\pder{}{h}\pder{}{t}
-h^{2}\pder[2]{}{h}  \right]\\
\frac{\partial^2 }{\partial u_3 \partial u_2} &=  \frac{t^{2}}{h^2(1-t^{2})}
\left[ -t (t-1)^{2} 
\pder[2]{}{t}-(1-t)^{2} \pder{}{t}-h(1-t)^{2}\pder{}{h}\pder{}{t}
 \right]
\end{split}
\]
follows that
\begin{equation}
\begin{split}
& \left[ t (t-1) 
\pder[2]{}{t}+(2t-1) \pder{}{t}-h\pder{}{h}\pder{}{t} \right] \log \tau
=  \omega_2 \bar{\omega}_2 \\
&\left[ t^{2} (t-1) 
\pder[2]{}{t}+t^{2} \pder{}{t}+h(t^{2}-2t)\pder{}{h}\pder{}{t}
-h^{2}\pder[2]{}{h}  \right] \log \tau = -\omega_3 \bar{\omega}_3\\
 &\left[ -t (t-1)^{2} 
\pder[2]{}{t}-(1-t)^{2} \pder{}{t}-h(1-t)^{2}\pder{}{h}\pder{}{t}
 \right] \log \tau = -\omega_1 \bar{\omega}_1\, .
\end{split}
\label{eq:ommm}
\end{equation}
Summing the right hand sides of \rf{ommm} and using \rf{Roms} we find
\begin{equation}
\omega_1 {\bar \omega}_1+ \omega_2 {\bar \omega}_2+ \omega_3 {\bar \omega}_3
= h^{2}\pder[2]{}{h} \log \tau  = -R^2\, ,
\label{eq:Romsf}
\end{equation}
which leads to a separation of variables formula
\begin{equation}
\log \tau  = R^{2} \log h +\log\tau_0 (t)\, ,
\label{eq:ff0}
\end{equation}
where on the right hand side we have 
introduced a single isomonodromic tau 
function $\tau_0$, which solely depends  on single variable
$t$. 
Let,
\[
f = t(t-1) \dder{\log \tau_0}{t} \, .
\]
Substituting relation \rf{ff0} into equation \rf{ommm} we arrive at
a following parametrization
of $\omega_i$'s in terms of $\tau_0$ or rather $f$:
\begin{equation}
\omega_2 {\bar \omega}_2= f^\prime, \;\;\; \omega_3 {\bar \omega}_3 = f-t f^\prime -R^{2}, \;\;\; 
\omega_1 {\bar \omega}_1=-f+(t-1)  f^\prime\, .
\label{eq:ftom}
\end{equation}
Taking a derivative of \rf{ftom} yields
\begin{equation}
t \dder{ \left(\omega_1 {\bar \omega}_1\right)}{t}= t(t-1)
\dder{ \left(\omega_2 {\bar \omega}_2\right)}{t}= -(t-1) 
\dder{ \left(\omega_3 {\bar \omega}_3\right)}{t}
= t (t-1) f^{\prime\prime} \, .
\label{eq:property}
\end{equation}
On the other hand from the generalized Euler top eqs. \rf{omeqs} it follows that
\begin{equation}
t \dder{ \left(\omega_1 {\bar \omega}_1\right)}{t}= t(t-1)
\dder{ \left(\omega_2 {\bar \omega}_2\right)}{t}= -(t-1) 
\dder{ \left(\omega_3 {\bar \omega}_3\right)}{t}
= {\bar \omega}_1\omega_2{\bar \omega}_3
+\omega_1{\bar \omega}_2\omega_3 \, .
 \label{eq:euta}
\end{equation}
Thus
\begin{equation}
t (t-1) f^{\prime\prime} ={\bar \omega}_1\omega_2{\bar
\omega}_3+\omega_1{\bar \omega}_2\omega_3 \, .
\label{eq:ttff}
\end{equation}
Take now the determinant of ${\bar V}+\nu$ $\colon$
\[
\rm{Det} \left({\bar V}+\underline{ \nu}\right)= \nu_1\nu_{2} \nu_{3} +{\bar \omega}_1\omega_2{\bar \omega}_3
 -\omega_1{\bar \omega}_2\omega_3 + \sum_{i=1}^3 \nu_{i} \omega_i {\bar
 \omega}_i = \rm{Det} \left(\underline{ \mu}\right)=  \mu_1\mu_{2} \mu_{3}
\]
or
\begin{equation}
{\bar \omega}_1\omega_2{\bar \omega}_3
 -\omega_1{\bar \omega}_2\omega_3 = \prod_{i=1}^{3} \mu_{i}-\prod_{i=1}^{3} \nu_{i}
  - \sum_{i=1}^3 \nu_{i} \omega_i {\bar \omega}_i \, .
\label{eq:omsboms}
\end{equation}
By squaring eqs. \rf{ttff} and \rf{omsboms} we obtain
\begin{equation}
\begin{split}
t^2(t-1)^2 \left(f^{\prime \prime}\right)^2 &= 4 \prod_{i=1}^{3} \omega_i {\bar \omega}_i
 + \prod_{i=1}^{3} \mu_{i}^{2}+\prod_{i=1}^{3} \nu_{i}^{2}-2  \prod_{i=1}^{3}
  \mu_{i} \nu_i +(\sum_{i=1}^3 \nu_{i} \omega_i {\bar \omega}_i)^{2}\\
  &-2 \left( \prod_{i=1}^{3} \mu_{i}-\prod_{i=1}^{3} \nu_{i}\right)
  \sum_{i=1}^3 \nu_{i}
  \omega_i {\bar \omega}_i\\
&=4  (-f+(t-1)f^{\prime })f^{\prime} (-R^{2}+f-tf^{\prime })\\
&+
  \left( \nu_{1}(-f+(t-1)f^{\prime })+\nu_{2}f^{\prime}+\nu_{3} (-R^{2}+f-tf^{\prime})\right)^{2}\\
&-2 \left(\prod_{i=1}^{3} \mu_{i}-\prod_{i=1}^{3} \nu_{i}\right) \left( \nu_{1}
  (-f+(t-1)f^{\prime })+\nu_{2}f^{\prime}+\nu_{3} (-R^{2}+f-tf^{\prime})\right)\\
&+\left(\prod_{i=1}^{3} \mu_{i}-\prod_{i=1}^{3} \nu_{i}\right)^{2} \\
&= -4 \, \bigg[\,
 f^{\prime}\, (tf^{\prime }-f)^2 - (f^{\prime})^2\, (tf^{\prime }-f)+c_5 \,(tf^{\prime }-f)^2 + c_6 \,f^{\prime}\,(tf^{\prime }-f)\\
&+c_7 \,(f^{\prime})^2+ c_8 \,(tf^{\prime }-f) +c_9\, f^{\prime} +c_{10}
\,\bigg] \, .
\end{split}
\label{eq:nonlineq} 
  \end{equation}
where coefficients :
\begin{equation}
\begin{split}
c_5 &= \left(- \frac{1}{4}\right) \, \left(\,{\nu_1}\,-\,{\nu_3}\,\right)^{2}\\
c_6 &= \left(- \frac{1}{4}\right) \,\left( -4\,{R}^{2}+ 2\left(
\,\nu_2\,-\,\nu_1\right)\,\left({\nu_1}\,-\,\nu_3\right)\,\right)\\
c_7 &= \left(- \frac{1}{4}\right) \,\left(4\,{R}^{2}+\left(\nu_1\,-\,\nu_2\right)^{2} \right)\\
c_8 &= \left(- \frac{1}{4}\right) \,\left(2\,\left(\nu_3-\nu_1\right)\right)\,
\left(\,\mu_1\,\mu_2\,\mu_3\,- \,\nu_1\nu_2\,\nu_3\,+\,\nu_3\,{R}^{2}\right)\\
c_9 &= \left(- \frac{1}{4}\right) \,\left(2\,\left(\nu_1-\nu_2\right)\right)\,
\left(\,\mu_1\,\mu_2\,\mu_3\,- \,\nu_1\nu_2\,\nu_3\,+\,\nu_3\,{R}^{2}\right)\\
c_{10} &= \left(- \frac{1}{4}\right) \left(\,\mu_1\,\mu_2\,\mu_3\,- 
\,\nu_1\nu_2\,\nu_3\,+\,\nu_3\,{R}^{2}\right)^2
\end{split}
\label{eq:cosgrov2} 
\end{equation}
were introduced to make it it more convenient to compare with
the technical steps taken in \cite{cosgrove} (see also \cite{conte}).
We perform a change variable $f \to \rho$ as follows. We set
\[ 
f = t(t-1) \frac{d \log \tau_0}{d t}= \rho + \,a \, t\, +\,b
\]
Thus $  f^{\prime\prime} = \rho^{\prime\prime}$,   $f^{\prime} = \rho^{\prime} + \,a$ and 
$t\,f^{\prime} -f = t \,\rho^{\prime}-\rho - \,b$.
Further we set 
\[
a = -c_5 , \qquad  b= c_5 + \h c_6
\]
in order to eliminate terms with $c_5,c_6$ in equation
\rf{nonlineq}.
This way we obtain for $\rho$:
\begin{equation}
\begin{split}
\left[t(t-1)\right]^2 \left(\rho^{\prime \prime}\right)^2 &= -4 \, \bigg[\,
 \rho^{\prime}\, (t\rho^{\prime }-\rho)^2 - (\rho^{\prime})^2\, (t\rho^{\prime
 }-\rho)\bigg]\\
&+A_1 \,(\rho^{\prime})^2+ A_2 \,(t\rho^{\prime }-\rho) +
A_3\, \rho^{\prime} +A_{4}\, ,
\end{split}
\label{eq:cosgrov3} 
\end{equation}
where
\begin{equation}
\begin{split}
A_1 &= (-4) \left( c_7 + c_5 + \h c_6\right)\\
A_2 &= (-4) \left( c_8 - c_5^2 -c_5 c_6\right)\\
A_3 &= (-4) \left( c_9 - c_5^2 - \frac{1}{4}c_6^2-c_5 c_6-2 c_5c_7\right)\\
A_4 &= (-4) \left( c_{10}+\frac{3}{2} c_5^2 c_6
+\frac{1}{2}c_5 c_6^2 + c_5^3 
+c_5^2 c_7-c_5 c_8- \h c_6 c_8-c_5 c_9 \right)\, .
\end{split}
\label{eq:cosgrov4} 
\end{equation}
Now we compare eq. \rf{nonlineq} to the Jimbo-Miwa-Okamoto 
$\sigma$-form of the Painlev\'e VI equation \rf{jmo}
with
\begin{alignat}{2}
v_1&= \h \left( \sqrt{-2\beta}+\sqrt{2\gamma}\right),& \qquad\quad
v_2= \h \left( \sqrt{-2\beta}-\sqrt{2\gamma}\right), \\
v_3&=\h \left( \sqrt{2\alpha}+\sqrt{1-2\delta}-1\right), &\quad
v_4=\h \left(- \sqrt{2\alpha}+\sqrt{1-2\delta}-1\right)
\end{alignat}

Expanding the products in \rf{jmo} and dividing by $\sigma^\prime$ 
gives 
\begin{equation}
\begin{split}
&t^2 (t-1)^2 (\sigma^{\prime\prime})^2 
+4 \, \bigg[\,
 \sigma^{\prime}\, (t\sigma^{\prime }-\sigma)^2 - (\sigma^{\prime})^2\, (t\sigma^{\prime
 }-\sigma)\bigg]
 -4  v_1v_2v_3v_4 \left(t \sigma^{\prime} -\sigma\right)\\
&=(\sigma^{\prime})^2 \left(\sum_{k=1}^4 v_k^2\right)
+\sigma^{\prime}\left(\sum_{i \ne j}^4 v_i^2 v_j^2-2  v_1v_2v_3v_4 \right)+\sum_{i \ne j\ne k} 
v_i^2 v_j^2 v_k^2
\label{eq:jmo1}
\end{split}
\end{equation}
Comparing with eq. \rf{cosgrov3} we see that the equations will 
agree for $\rho=\sigma$ and :
\[
\begin{split}
  A_{1}&=  \sum_{k=1}^4 v_k^2 , \qquad \qquad\qquad\qquad\qquad
  A_{2} = 4  v_1v_2v_3v_4 \\
  A_{3} &= \left(\sum_{i \ne j}^4 v_i^2 v_j^2-2  v_1v_2v_3v_4 \right)
  , \qquad A_{4} =  \sum_{i \ne j\ne k} 
v_i^2 v_j^2 v_k^2 \, ,
\end{split}
\]
with $v_{k}^{2}, k=1,2,3,4$ being roots
of the fourth-order polynomial
$x^{4} -A_1 x^{3} + \left(A_3+  A_2/2\right) x^{2} - A_4 x +  A_2^{2}/16$.
Plugging the values of $A_i$'s from equation \rf{cosgrov4} 
into the above fourth-order polynomial
leads to the following generic solution for its roots :
\begin{equation}
v^2_i=  \left( \frac{\nu_1+\nu_3}{2} -\mu_i\right)^2, \; i=1,2,3,
 \qquad\;\;
v^2_4=  \left(\frac{\nu_1-\nu_3}{2}\right)^2  \, .
\label{eq:x1234}
\end{equation}
To satisfy condition $A_{2} = 4  v_1v_2v_3v_4 $ we choose 
\begin{equation}
v_i=   \frac{\nu_1+\nu_3}{2} -\mu_i, \; i=1,2,3,
 \qquad\;\;
v_4=  \frac{\nu_1-\nu_3}{2}
\label{eq:posvis}
\end{equation}
and comparing with relation \rf{via} we get 
(c.f. \cite{Boalch}):
\begin{alignat}{2} 
\b&=-\frac{(\nu_2-\mu_i)^2}{2},&\qquad 
\g=\frac{(\mu_j-\mu_k)^2}{2} \nonumber  \\
\a&=\frac{(\nu_3-\mu_i)^2}{2}
,&\qquad
\d= \frac{1}{2} \left(1 -\left( 1+\nu_1-\mu_i\right)^2\right)\, ,
\label{eq:param1}
\end{alignat}
with $(i,j,k)=(3,1,2)$.
Due to manifest $D_4$ symmetry of the Painlev\'e VI equation 
\rf{jmo} any permutation of $v_1,v_2,v_3,v_4$ 
as well as change of signs in front of even number of $v_i$'s in 
equation \rf{via} will lead to other equivalent solutions.

For example for solution 
\begin{equation}
v_i= (-1)^i \left( \frac{\nu_1+\nu_3}{2} -\mu_i\right), \; i=1,2,3,
 \qquad\;\;
v_4=  \frac{\nu_1-\nu_3}{2} \, ,
\label{eq:negvis}
\end{equation}
obtained from \rf{posvis} by simultaneously changing signs
in front of $v_1$ and $v_3$ we obtain 
\begin{alignat}{2} 
\beta&= - {\displaystyle \frac {(\mu_1 - \mu_2)^{2}}{2}}
,&\qquad 
\gamma=  {\displaystyle \frac {(\nu_2 - \mu_3)^{2}}{2}}, \nonumber  \\
 \alpha  &=  \frac{(\nu_1- \mu_3)^2}{2} ,&\qquad 
\delta= \h \left[1- \left(1+\mu_3-\nu_3\right)^2\right]
\, .
\label{eq:param2}
\end{alignat}
Furthermore, exchanging $v_1 \leftrightarrow v_3$ in equation \rf{posvis}
and next exchanging $v_2 \leftrightarrow v_3$ in equation \rf{posvis}
we get relation \rf{param1} with $(i,j,k)=(1,2,3)$ and 
$(i,j,k)=(2,1,3)$, respectively.

The relation \rf{param1} compares well with 
the result found in 
\cite{kakei} after inserting $\mu_3=-1$, needed 
in \cite{kakei} in order to establish equivalence with
the system of $2 \times 2$ Schlesinger equations.

Note, that the relation \rf{rh-numu} is invariant under
simultaneous transformations
$\underline{ \nu} \to  \underline{ \nu}+c I$ and
$\underline{ \mu} \to  \underline{ \mu}+c I$,
with a constant $c$. 
It is indeed easy to confirm by checking all the above expressions 
for the coefficients $v_i, i=1,{\ldots} ,4$ that they
are left invariant under
$ \nu_i \to  \nu_i + c,\; \mu_i \to  \mu_i + c$ for $i=1,2,3$.
This fact will be used later below eq. (\ref{e109}).

$A_{i}, i=1,\dots,4$ can also be expressed directly in terms 
of the Painlev\'e parameters $\a, \b,\g, \d$ via relation \rf{via} 
as follows 
\begin{equation}
\begin{split}
A_{1}&= -\beta  + \gamma + \alpha  - \delta  -
\sqrt{1 - 2\,\delta } +1\\
A_{2}&=  \left(\beta  + \gamma\right)\left( \alpha +\delta +\sqrt{1 - 2\,\delta }
-1\right)\\
A_{3}&= \left( \b -\g \right) \left(- \a +\d +\sqrt{1-2\d}-1\right)+\frac{1}{4}
\left(- \a-\d -\sqrt{1-2\d}+\b + \g+1 \right)^{2} \\
A_{4}&= -\frac{1}{4}  \left( \b - \g \right) \left( \a +\d  +\sqrt{1-2\d}-1\right)^2
+\frac{1}{4} \left( \b + \g \right)^{2}	\left( \a  -\d-\sqrt{1-2\d}+1
\right)\, .
\end{split}
\label{eq:AApa}
\end{equation}
According to \cite{Okamoto} the solution $y$ 
to the Painlev\'e VI equation \rf{P6} is obtained from
$\sigma$ via relation:
\begin{equation}
y=\frac{1}{2A}\left((v_3+v_4)B+\left(\frac{d\sigma}{dt}-v_3v_4
\right)C\right)\, ,
\label{eq:okay}
\end{equation}
where
\begin{equation}
A=\left(\frac{d\sigma}{dt}+v_3^2
\right)\left(\frac{d\sigma}{dt}+v_4^2
\right)\, ,
\end{equation}
\begin{equation}
B=t(t-1)\frac{d^2\sigma}{dt^2}+(v_1+v_2+v_3+v_4)\frac{d\sigma}{dt}-(v_1v_2v_3+v_1v_2v_4+v_1v_3v_4+v_2v_3v_4)
\end{equation}
and
\begin{equation}
C=2\left(t\frac{d\sigma}{dt}-\sigma
\right)-(v_1v_2+v_1v_3+v_1v_4+v_2v_3+v_2v_4+v_3v_4)\, .
\end{equation}

\section{Construction of the tau function}

In this section we will construct the tau function discussed
above. 
We will
follow \cite{KL}, and introduce a semi-infinite wedge space $F =
\Lambda^{\frac{1}{2}\infty} {\mathbb C}^{\infty}$ as the vector space
with a basis consisting of all semi-infinite monomials of the form
$v_{{\mathfrak i}_{1}} \wedge v_{{\mathfrak i}_{2}} \wedge v_{{\mathfrak
i}_{3}} \ldots$,
with ${\mathfrak i}_j\in
\frac{1}{2}+\mathbb{Z}$, where ${\mathfrak i}_{1} >
{\mathfrak i}_{2} > {\mathfrak i}_{3} > \ldots$ and ${\mathfrak i}_{\ell +1} =
{\mathfrak i}_{\ell} -1$ for $\ell >>
0$.  Define the wedging and contracting
operators $\psi^{+}_{\mathfrak j}$ and $\psi^{-}_{\mathfrak j}\ \ ({\mathfrak
j} \in {\mathbb Z} +
\frac{1}{2})$ on $F$ by
\[
\begin{aligned}
&\psi^{+}_{\mathfrak j} (v_{{\mathfrak i}_{1}} \wedge v_{{\mathfrak i}_{2}}
\wedge \cdots ) = \begin{cases} 0
& \text{if}\ -{\mathfrak j} = {\mathfrak i}_{s}\ \text{for some}\ s \\
(-1)^{s} v_{{\mathfrak i}_{1}} \wedge v_{{\mathfrak i}_{2}}
\cdots \wedge v_{{\mathfrak i}_{s}} \wedge
v_{-{\mathfrak j}} \wedge v_{{\mathfrak i}_{s+1}}
\wedge \cdots &\text{if}\ {\mathfrak i}_{s} > -{\mathfrak j} >
{\mathfrak i}_{s+1}\end{cases} \\
&\psi^{-}_{\mathfrak j} (v_{{\mathfrak i}_{1}} \wedge v_{{\mathfrak i}_{2}}
\wedge \cdots ) = \begin{cases} 0
&\text{if}\ {\mathfrak j} \neq {\mathfrak i}_{s}\ \text{for all}\ s \\
(-1)^{s+1} v_{{\mathfrak i}_{1}} \wedge v_{{\mathfrak i}_{2}} \wedge \cdots
\wedge
v_{{\mathfrak i}_{s-1}} \wedge v_{{\mathfrak i}_{s+1}} \wedge
\cdots &\text{if}\ {\mathfrak j} = {\mathfrak i}_{s}.
\end{cases}
\end{aligned}
\]
These operators satisfy the following relations
$({\mathfrak i},{\mathfrak j} \in {\mathbb Z}+\frac{1}{2}, \lambda ,\mu =
+,-)$:
\begin{equation}
\label{comm}\psi^{\lambda}_{\mathfrak i} \psi^{\mu}_{\mathfrak j} +
\psi^{\mu}_{\mathfrak j}
\psi^{\lambda}_{\mathfrak i} = \delta_{\lambda ,-\mu} \delta_{{\mathfrak
i},-{\mathfrak j}},
\end{equation}
hence they generate a Clifford algebra, which we denote by ${\cal C}\ell$.

Introduce the following elements of $F$ $(m \in {\mathbb Z})$:
\[
|m\rangle = v_{m-\frac{1}{2} } \wedge v_{m-\frac{3}{2} } \wedge
v_{m-\frac{5}{2} } \wedge \cdots .\]
It is clear that $F$ is an irreducible ${\cal C}\ell$-module such that
\[
\psi^{\pm}_{\mathfrak j} |0\rangle = 0 \ \text{for}\ {\mathfrak j} > 0 .
\]
Think of the adjoint module $F^*$ in the following way,
it
is the vector space
with a basis consisting of all semi-infinite monomials of the form
$\ldots \wedge v_{{\mathfrak i}_{3}} \wedge v_{{\mathfrak i}_{2}} \wedge
v_{{\mathfrak i}_{1}} $,
where ${\mathfrak i}_{1} <
{\mathfrak i}_{2} < {\mathfrak i}_{3} < \ldots$ and ${\mathfrak i}_{\ell +1} =
{\mathfrak i}_{\ell} +1$ for $\ell >>
0$. The
operators $\psi^{+}_{\mathfrak j}$ and $\psi^{-}_{\mathfrak j}\ \ ({\mathfrak
j} \in {\mathbb Z} +
\frac{1}{2})$ also act on $F^*$ by contracting and wedging, but in a different
way, viz.,
\[\begin{aligned}
&(\cdots\wedge v_{{\mathfrak i}_{2}} \wedge v_{{\mathfrak i}_{1}}
)\psi^{+}_{\mathfrak j}  = \begin{cases} 0
& \text{if}\ {\mathfrak j}\ne {\mathfrak i}_{s}\ \text{for all}\ s \\
(-1)^{s+1} \cdots\wedge v_{{\mathfrak i}_{s+1}}\wedge v_{{\mathfrak i}_{s-1}}
\wedge \cdots
v_{{\mathfrak i}_{2}} \wedge v_{{\mathfrak i}_{1}}&\text{if}\ {\mathfrak i}_{s}
= {\mathfrak j}\end{cases} \\
& (\cdots\wedge v_{{\mathfrak i}_{2}} \wedge v_{{\mathfrak i}_{1}}
)\psi^{-}_{\mathfrak j} = \begin{cases} 0
&\text{if}
\  -{\mathfrak j} = {\mathfrak i}_{s}\ \text{for some}\ s \\
(-1)^{s} \cdots\wedge v_{{\mathfrak i}_{s+1}}\wedge v_{\mathfrak j}\wedge
v_{{\mathfrak i}_{s}} \wedge \cdots
v_{{\mathfrak i}_{2}} \wedge v_{{\mathfrak i}_{1}}&\text{if}\ {\mathfrak
i}_{s}<-{\mathfrak j}<{\mathfrak i}_{s+1}.
\end{cases}
\end{aligned}
\]
We introduce the element $\langle m | $, by
\[
\langle m | = \cdots \wedge v_{m+\frac{5}{2} } \wedge v_{m+\frac{3}{2} }
\wedge
v_{m+\frac{1}{2} } \, ,
\]
such that
\[
\langle 0 |\psi^{\pm}_{\mathfrak j}  = 0 \ \text{for}\ {\mathfrak j} < 0\, .
\]
We define the vacuum expectation value by
\[
\langle 0 |0\rangle=1,\qquad\mbox{ and denote }\quad \langle A\rangle=
\langle
0 |A|0\rangle\, .
\]

We next identify $\mathbb{C}^\infty$ with the space 
$\mathbb{C}[z, z^{-1}]^N$ as follows.
Let $\{ e_i\}_{1\le i\le N}$ be a basis of $\mathbb{C}^N$,
we relabel the basis vectors $v_{\mathfrak i}$ 
\begin{equation}
\label{relabelv}
v^{({j})}_{\mathfrak k} =z^{-{\mathfrak k}-\frac12}e_j= v_{N{\mathfrak k} - \frac{1}{2}(N-2j+1)},
\end{equation}
and with them the
corresponding fermionic operators (the wedging and contracting operators):
\[
\psi^{\pm (j)}_{\mathfrak k} = \psi^{\pm}_{N{\mathfrak k}
\pm\frac{1}{2}(N-2j+1)}.
\]
Note that with this relabeling we have:
\[
\psi^{\pm (j)}_{\mathfrak k}|0\rangle = 0\ \text{for}\ {\mathfrak k} > 0
\]
and
\[
| 0\rangle=z^0e_N\wedge z^0e_{N-1}\wedge\cdots
\wedge z^0e_{1}\wedge ze_N\wedge ze_{N-1}\wedge\cdots
\wedge ze_{1}\wedge z^2e_N\wedge z^2e_{N-1}\wedge\cdots\, .
\]

Introduce the  fermionic fields $(0\ne \lambda \in {\mathbb C})$:
\[
\psi^{\pm (j)}(\lambda)=\sum_{{\mathfrak k} \in {\mathbb
Z}+\frac{1}{2}} \psi^{\pm
(j)}_{\mathfrak k} \lambda^{-{\mathfrak k}-\frac{1}{2}}.
\]
Next, we introduce bosonic fields $(1 \leq i \leq n)$:
\[
\alpha^{(i)}(\lambda) \equiv \sum_{k \in {\mathbb Z}} \alpha^{(i)}_{ k}
\lambda^{-{k}-1}
= :\psi^{+(i)}(\lambda) \psi^{-(i)}(\lambda):, 
\]
where $:\ :$ stands for the {\it normal ordered product} defined in
the usual way $(\lambda ,\mu = +$ or $-$):
\[
:\psi^{\lambda (i)}_{\mathfrak k} \psi^{\mu (j)}_{\mathfrak l}: =
\begin{cases} \psi^{
\lambda (i)}_{\mathfrak k}
\psi^{\mu (j)}_{\mathfrak l}\ &\text{if}\ {\mathfrak l} > 0 \\
-\psi^{\mu (j)}_{\mathfrak l} \psi^{\lambda (i)}_{\mathfrak k} &\text{if}\
{\mathfrak l} <
0.\end{cases} 
\]
One checks (using e.g. the Wick formula) that the operators
$\alpha^{(i)}_{k}$ satisfy the canonical commutation relation of the associative
oscillator algebra,  which we
denote by ${\mathfrak a}$:
\[
{}[\alpha^{(i)}_{k},\alpha^{(j)}_{\ell}] =
k\delta_{ij}\delta_{k,-\ell},
\]
and one has
\[
\alpha^{(i)}_{k}|m\rangle = 0 \ \text{for}\ k > 0,\qquad
\langle m|\alpha^{(i)}_{k}= 0 \ \text{for}\ k < 0.
\]
In order to express the fermionic fields $\psi^{\pm (i)}(\lambda)$ in terms of
the bosonic fields $\alpha^{(i)}(\lambda)$, we need some additional operators
$Q_{i},\ i = 1,2,\ldots, N$, on $F$.  These operators are uniquely defined by
the following conditions:
\begin{equation}
\label{Q}
Q_{i}|0\rangle = \psi^{+(i)}_{-\frac{1}{2}} |0\rangle ,\ Q_{i}\psi^{\pm
(j)}_{\mathfrak k} = (-1)^{\delta_{ij}+1} \psi^{\pm
(j)}_{{\mathfrak k}\mp \delta_{ij}}Q_{i}.
\end{equation}
They satisfy the following commutation relations:
\[
Q_{i}Q_{j} = -Q_{j}Q_{i}\ \text{if}\ i \neq j,\ [\alpha^{(i)}_{k},Q_{j}] =
\delta_{ij} \delta_{k0}Q_{j}.
\]

We shall use below the following notation
\[
|k_{1},k_{2}, \ldots , k_N\rangle = Q^{k_{1}}_{1}
Q^{k_{2}}_{2}\cdots Q^{k_{N}}_{N}|0\rangle ,\qquad
\langle k_{1},k_{2}, \ldots , k_N|=\langle 0|Q^{-k_{N}}_{N}\cdots
Q^{-k_{2}}_{2}Q^{-k_{1}}_{1},
 \]
 such that
 \[
 \langle k_{1},k_{2}, \ldots ,k_N|k_{1},k_{2}, \ldots , k_N\rangle=\langle
0|0\rangle=1
 \, .\]
One easily checks the following relations:
\[
{}[\alpha^{(i)}_{k},\psi^{\pm (j)}_{\mathfrak m}] = \pm \delta_{ij} \psi^{\pm
(j)}_{k+{\mathfrak m}} 
\]
and
\[
\begin{split}
Q_i^{\pm 1}|k_{1},k_{2},\ldots , k_N\rangle
=&(-)^{k_1+k_2+\cdots+k_{i-1}}|k_{1},k_{2},\ldots ,k_{i-1},k_i\pm 1,k_{i+1},
\ldots ,
k_N\rangle ,\\
\langle k_{1},k_{2}, \ldots ,k_N|Q_i^{\pm 1}=&
(-)^{k_1+k_2+\cdots+k_{i-1}}\langle k_{1},k_{2},\ldots ,k_{i-1},k_i\mp
1,k_{i+1}
,\ldots ,
k_N|\, .
\end{split}
\]
These formulas lead to the following vertex operator expression for
$\psi^{\pm
(i)}(\lambda)$.
Given any sequence $s=(s_1,s_2,\dots)$, define
\[
\Gamma^{(j)}_\pm( s)=\exp \left(\sum_{k=1}^\infty s_k \alpha^{(j)}_{\pm
k}\right),
\]
then (see e.g. 
\cite{DJKM1}, \cite{TV})
\[
\begin{aligned}\psi^{\pm (i)}(\lambda) &= Q^{\pm 1}_{i}\lambda^{\pm \alpha^{(i)}_{0}} \exp
(\mp \sum_{k < 0} \frac{1}{k} \alpha^{(i)}_{k}\lambda^{-k})\exp(\mp
\sum_{k > 0} \frac{1}{k} \alpha^{(i)}_{k} \lambda^{-k})\\[3mm]
&=Q^{\pm 1}_{i}\lambda^{\pm \alpha^{(i)}_{0}}\Gamma_{-}^{(i)}(\pm
[\lambda])\Gamma_{+}^{(i)}(\mp
[\lambda^{-1}]),
\end{aligned}
 \]
 where $[\lambda]=(\lambda,\frac{\lambda^2}{2}, \frac{\lambda^3}{3},\ldots)$.
Note, that
\begin{equation}\label{Gamma_vac}
\begin{split}
\Gamma_+^{(j)}(s) \,|k_{1},k_{2}, \ldots ,k_n\rangle
=& |k_{1},k_{2}, \ldots ,k_n\rangle  \,,\\
\langle k_{1},k_{2}, \ldots , k_n| \, \Gamma_-^{(j)}(s)
=& \langle k_{1},k_{2}, \ldots ,k_n|  \,.
\end{split}
\end{equation}
Also observe that 
\begin{equation}\label{Gamma_Gamma}
\Gamma_+^{(j)}(s)\, \Gamma_-^{(k)}(s') = \gamma(s,s')^{\delta_{jk}}\,
\Gamma_-^{(k)}(s')\, \Gamma_+^{(j)}(s) \, ,
\end{equation}
where
\[
\gamma(s,s')=e^{\sum n\, s_n s'_n} \, .
\]
We have
\begin{align}
\Gamma_\pm^{(j)}(s) \, \psi^{+(k)}(\lambda)  &= \gamma(s,[\lambda^{\pm 1}])^{\delta_{jk}}
\, \psi^{+(k)}(\lambda) \, \Gamma_\pm^{(j)}(s) \,, \notag  \\
\Gamma_\pm^{(j)}(s) \, \psi^{-(k)}(\lambda)  &= \gamma(s,-[\lambda^{\pm 1}])^{\delta_{jk}}
\, \psi^{-(k)}(\lambda) \, \Gamma_\pm^{(j)} (s)\,. \label{e111}
\end{align}
Note that
\begin{equation}\label{e109}
\gamma(t,[\lambda])= \exp\left(\sum_{n\ge 1} t_n\, \lambda^n\right) \, .
\end{equation}

Assume from now on that $N=3$. In the 3-component setting we will
construct  the element $g_0 z^{-\underline\mu}$ (cf. (\ref{eq:gz-form}) 
acting on the vacuum $|0\rangle$.
It is obvious, see e.g. \cite{KL}, that this element in the orbit will satisfy the 
3-component KP-hierarchy.
To obtain the most general construction, we do not assume any order in the size of 
the $\mu_i$. But we do the following. 
Since 
\rf{rh-numu} is invariant under
simultaneous transformations
$\underline{ \nu} \to  \underline{ \nu}+c I$ and
$\underline{ \mu} \to  \underline{ \mu}+c I$,
with a constant $c$, we subtract
such a constant $c$ from all $\mu_i$  and $\nu_i$ 
such that  all $\mu_i$ become non-positive.
Define
\[
 m_1=\max\{ -\mu_i\}
\]
and choose $m_2$ and $m_3$ in such a way that
\[
m_1\ge m_2\ge m_3\ge 0\quad\mbox{and}\quad 
\{ m_1,m_2,m_3\} = \{ -\mu_1,-\mu_2,-\mu_3\}\, .
\]
In fact one can choose $c$ in such a way, without loss of generality, that  $m_3=0$, 
however we will not do that.

We start with an example. Take $-\mu_1=m_1=3$, $-\mu_3=m_2=2$ and $-\mu_2=m_3=0$ 
and note that the vacuum is equal to
\[
|0\rangle=e_3\wedge e_2\wedge e_1\wedge ze_3\wedge ze_2\wedge ze_1\wedge z^2e_3
\wedge z^2e_2\wedge z^2e_1\wedge z^3e_3\wedge z^3e_2\wedge z^3e_1\wedge z^4e_3\wedge \cdots
\]
Now let $G$ act on this.
Set
\[
g_0(e_1)=w^{(1)},\qquad g_0(e_2)=w^{(3)},\qquad g_0(e_3)=w^{(2)},
\]
thus $G=g_0 z^{-\underline\mu}$ acting on the vacuum gives
\[
G|0\rangle=z^2w^{(2)}\wedge w^{(3)}\wedge z^3w^{(1)}\wedge z^3w^{(2)}\wedge zw^{(3)}\wedge z^4w^{(1)}\wedge z^4w^{(2)}\wedge z^2w^{(3)}\wedge z^5w^{(1)}\wedge z^5w^{(2)}\wedge z^3w^{(3)}\wedge z^6w^{(1)}\wedge \cdots
\]
Now permute the elements infinitely many times, 
this gives up to a sign,
\[
G|0\rangle=\pm w^{(3)}\wedge zw^{(3)}\wedge z^2w^{(3)}\wedge z^2w^{(2)}\wedge  z^3w^{(1)}\wedge z^3w^{(2)}\wedge 
 z^3w^{(3)}\wedge z^4w^{(1)}\wedge z^4w^{(2)}\wedge  z^4w^{(3)}\wedge  \cdots
\]
Up to some multiplicative scalar $K$, which might be an infinite
constant, this is equal to
\[
\begin{split}
G|0\rangle=&K w^{(3)}\wedge zw^{(3)}\wedge z^2w^{(3)}\wedge z^2w^{(2)}\wedge  
z^3e_3\wedge z^3e_2\wedge 
 z^3e_1\wedge z^4e_3\wedge z^4e_2\wedge  z^4e_1\wedge  \cdots\\[3mm]
 =&L 
 \phi^{+(3)}_{\frac12}\phi^{+(3)}_{\frac32}\phi^{+(3)}_{\frac52}
 \phi^{+(2)}_{\frac52} \left( z^3e_3\wedge z^3e_2\wedge 
 z^3e_1\wedge z^4e_3\wedge z^4e_2\wedge  z^4e_1\wedge  \cdots\right)\\[3mm]
 =&M 
 \phi^{+(3)}_{\frac12}\phi^{+(3)}_{\frac32}\phi^{+(3)}_{\frac52}
 \phi^{+(2)}_{\frac52}|-3,-3,-3\rangle\, ,
 \end{split}
\]
where 
\[
\phi_{\mathfrak k}^{+(i)}=w_1^{(i)}\psi_{\mathfrak k}^{+(1)}+
w_2^{(i)}\psi_{\mathfrak k}^{+(2)}+w_3^{(i)}\psi_{\mathfrak k}^{+(3)},\qquad {\mathfrak k}\in \frac12 +\mathbb{Z}
\]
and 
\[
w^{(i)}=\left(w_1^{(i)},w_2^{(i)},w_3^{(i)}\right),\qquad i=1,2,3.
\]
Since we want to avoid infinite constants, we do not consider the element
$g_0 z^{-\underline\mu}| 0\rangle$, but rather 
\begin{equation}
\label{1}
G|0\rangle=\phi_{m_3+\frac12}^{+(3)}\phi_{m_3+\frac32}^{+(3)}\cdots
\phi_{m_1-\frac12}^{+(3)}
\phi_{m_2+\frac12}^{+(2)}
\phi_{m_2+\frac32}^{+(2)}\cdots
\phi_{m_1-\frac12}^{+(2)}| -m_1,-m_1,-m_1\rangle\, ,
\end{equation}
Let $(\nu_1,\nu_2,\nu_3)\in \mathbb{Z}^3$, 
we decompose the element $G|0\rangle$ as
\begin{equation}
\label{1a}
G|0\rangle=\sum_{\nu_1, 
\nu_2, 
\nu_3\in \mathbb{Z}}g(
\nu_1, 
\nu_2, 
\nu_3)|
\nu_1, 
\nu_2, 
\nu_3\rangle\, ,
\end{equation}
then
\[
g(
\nu_1, 
\nu_2, 
\nu_3)=\langle\nu_1, 
\nu_2, 
\nu_3|
G|0\rangle\, .
\]
Using the boson-fermion correspondence $\sigma$ as  in e.g. 
\cite{Kac-Raina}, \cite{KL} or \cite{vdLeur-Martini}, we can express
$g(
\nu_1, 
\nu_2, 
\nu_3)$ in terms of of the $u_i^{(k)}$, $i=1,2,3,\dots$, $k=1,2,3$ viz.
\begin{equation}
\label{1b}
\sigma(G|0\rangle)=\sum_{\nu_1, 
\nu_2, 
\nu_3\in \mathbb{Z}}\tau(
\nu_1,
\nu_2,
\nu_3;{\bf u})|\nu_1, 
\nu_2, 
\nu_3\rangle\, .
\end{equation}
We use the notation $u^{(k)}$ for all $u_i^{(k)}$, $i=1,2,3,\ldots$, 
and $\bf u$ for all $u_i^{(k)}$, $i=1,2,3,\ldots$,
$k=1,2,3$. Note, that we do not put all higher flows $u^{(k)}_i,\, i>1$ 
to zero yet. Now any $\tau(
\nu_1,
\nu_2,
\nu_3;{\bf u})$, with  all higher times $u^{(k)}_i,\, i>1$ set to zero, will turn out to be a tau function of the Painlev\'e VI equation.

One has 
\[
\tau(
\nu_1,
\nu_2,
\nu_3;{\bf u})=\langle\nu_1, 
\nu_2, 
\nu_3|e^{\sum_{i,k}\alpha^{(i)}u_k^{(i)}}
G|0\rangle=
\langle 0|Q_3^{-\nu_3}Q_2^{-\nu_2}Q_1^{-\nu_1}e^{\sum_{i,k}\alpha^{(i)}u_k^{(i)}}
G|0\rangle
\, .
\]
In this $3$-component Clifford algebra setting 
the element
$
Q_3^{-\nu_3}Q_2^{-\nu_2}Q_1^{-\nu_1}
$ is related to
$
z^{\underline \nu}
$ and
$e^{\sum_{i,k}\alpha_k^{(i)}u_k^{(i)}}$ to
$e^{\sum_{i,k}z^iE_{kk}u_k^{(i)}}$.

First note that 
\[
e^{\sum_{i,k}\alpha_k^{(i)}u_k^{(i)}}Q_3^{-\nu_3}Q_2^{-\nu_2}Q_1^{-\nu_1}
=Q_3^{-\nu_3}Q_2^{-\nu_2}Q_1^{-\nu_1}e^{\sum_{i,k}\alpha_k^{(i)}u_k^{(i)}}
\]
and that
\[
e^{\sum_{i,k}\alpha_k^{(i)}u_k^{(i)}}\psi_k^{+(i)}e^{-\sum_{i,k}\alpha_k^{(i)}u_k^{(i)}}
=\sum_{j=0}^\infty \psi_{k+j}^{+(i)}S_j(u^{(i)})\, 
\]
here the $S_j(t)$ are the elementary Schur functions.
Also
\[
e^{\sum_{i,k}\alpha_k^{(i)}u_k^{(i)}}| -m_1,-m_1,-m_1\rangle=| -m_1,-m_1,-m_1\rangle\, .
\]
Next, let 
$e^{\sum_{i,k}\alpha_k^{(i)}u_k^{(i)}}$ act on the element (\ref{1}). This gives
\begin{equation}
\label{2}
\begin{split}
&\tilde G({\bf u})|0\rangle:=\\[3mm]
&e^{\sum_{i,k}\alpha_k^{(i)}u_k^{(i)}}
\phi_{m_3+\frac12}^{+(3)}\phi_{m_3+\frac32}^{+(3)}\cdots
\phi_{m_1-\frac12}^{+(3)}
\phi_{m_2+\frac12}^{+(2)}
\phi_{m_2+\frac32}^{+(2)}\cdots
\phi_{m_1-\frac12}^{+(2)}| -m_1,-m_1,-m_1\rangle=\\[3mm]
&\quad
\phi_{m_3+\frac12}^{+(3)}({\bf u})\phi_{m_3+\frac32}^{+(3)({\bf u})}\cdots
\phi_{m_1-\frac12}^{+(3)}({\bf u})
\phi_{m_2+\frac12}^{+(2)}({\bf u})
\phi_{m_2+\frac32}^{+(2)}({\bf u})\cdots
\phi_{m_1-\frac12}^{+(2)}({\bf u})| -m_1,-m_1,-m_1\rangle\, ,
\end{split}
\end{equation}
where
\begin{equation}
\label{3}
\phi_k^{+(i)}({\bf u})=\sum_{a=1}^3
\sum_{j=k}^{m_1-\frac12} w_a^{(i)}\psi_{j}^{+(a)}S_{j-k}(u^{(a)})\, . 
\end{equation}

Next we let $Q_3^{-\nu_3}Q_2^{-\nu_2}Q_1^{-\nu_1}$ act on the element (\ref{2}). This gives 
\begin{equation}
\label{4}
\begin{split}
\hat G(\underline\nu;{\bf u})|0\rangle :=&
Q_3^{-\nu_3}Q_2^{-\nu_2}Q_1^{-\nu_1}
\phi_{m_3+\frac12}^{+(3)}({\bf u})\phi_{m_3+\frac32}^{+(3)({\bf u})}\cdots\\[3mm]
&\quad\cdots
\phi_{m_1-\frac12}^{+(3)}({\bf u})
\phi_{m_2+\frac12}^{+(2)}({\bf u})
\phi_{m_2+\frac32}^{+(2)}({\bf u})\cdots
\phi_{m_1-\frac12}^{+(2)}({\bf u})| -m_1,-m_1,-m_1\rangle\, 
\end{split}
\end{equation}
Then
\[
\begin{split}
\tau&({\underline \nu};{\bf u})=\tau(
\nu_1,
\nu_2,
\nu_3;{\bf u})=\langle 0|\hat G(\underline\nu;{\bf u})|0\rangle=\langle 0|Q_3^{-\nu_3}Q_2^{-\nu_2}Q_1^{-\nu_1}\times\\[3mm]
&\quad\phi_{m_3+\frac12}^{+(3)}({\bf u})\phi_{m_3+\frac32}^{+(3)({\bf u})}\cdots
\phi_{m_1-\frac12}^{+(3)}({\bf u})
\phi_{m_2+\frac12}^{+(2)}({\bf u})
\phi_{m_2+\frac32}^{+(2)}({\bf u})\cdots
\phi_{m_1-\frac12}^{+(2)}({\bf u})| -m_1,-m_1,-m_1\rangle\, ,
\end{split}
\]
Define 
\[
f_a^{(b)}({\underline \nu})=(-)^{\nu_1+\nu_2+\nu_3+\nu_a}w_a^{(b)}
\]
then
\begin{equation}
\label{5}
\begin{split}
\hat \phi_\ell^{+(b)}({\underline \nu};{\bf u}):=&
Q_3^{-\nu_3}Q_2^{-\nu_2}Q_1^{-\nu_1}\phi_\ell^{+(b)}({\bf u})
Q_1^{\nu_1}Q_2^{\nu_2}Q_3^{\nu_3}\\[3mm]
=&\sum_{a=1}^3
\sum_{j=\ell}^{m_1-\frac12} (-)^{\nu_1+\nu_2+\nu_3+\nu_a}
w_a^{(b)}\psi_{j+\nu_a}^{+(a)}S_{j-\ell}(u^{(a)})\\[3mm]
=&\sum_{a=1}^3\sum_{j=\ell}^{m_1-\frac12} f_a^{(b)}({\underline 
\nu})\psi_{j+\nu_a}^{+(a)}S_{j-\ell}(u^{(a)})\\[3mm]
=&\sum_{a=1}^3\sum_{j=\ell+\nu_a}^{m_1+\nu_a-\frac12} f_a^{(b)}({\underline \nu})\psi_{j}^{+(a)}S_{j-\ell-\nu_a}(u^{(a)})
\end{split}
\end{equation}
and
\[
Q_3^{-\nu_3}Q_2^{-\nu_2}Q_1^{-\nu_1}| -m_1,-m_1,-m_1\rangle
=(-)^{m_1\nu_2+\nu_1\nu_2+\nu_1\nu_3+\nu_2\nu_3}| -m_1-\nu_1,-m_1-\nu_2,-m_1-\nu_3\rangle\, .
\]
Thus
\begin{equation}
\label{6aa}
\begin{split}
\hat G&({\underline \nu};{\bf u})|0\rangle=(-)^{m_1\nu_2+\nu_1\nu_2+\nu_1\nu_3+\nu_2\nu_3}
\hat\phi_{m_3+\frac12}^{+(3)}({\underline \nu};{\bf u})\hat\phi_{m_3+\frac32}^{+
(3)({\underline \nu};{\bf u})}\cdots \hat\phi_{m_1-\frac12}^{+(3)}({\underline \nu};{\bf u})
\times\\[3mm]
&\hat\phi_{m_2+\frac12}^{+(2)}({\underline \nu};{\bf u})
\hat\phi_{m_2+\frac32}^{+(2)}({\underline \nu};{\bf u})\cdots
\hat\phi_{m_1-\frac12}^{+(2)}({\underline \nu};{\bf u})| -m_1-\nu_1,-m_1-\nu_2,-m_1-
\nu_3\rangle
\end{split}
\end{equation}
and
\begin{equation}
\label{6}
\begin{split}
\tau&({\underline \nu};{\bf u})=(-)^{m_1\nu_2+\nu_1\nu_2+\nu_1\nu_3+\nu_2\nu_3}
\langle 0|\hat\phi_{m_3+\frac12}^{+(3)}({\underline \nu};{\bf u})
\hat\phi_{m_3+\frac32}^{+(3)}({\underline \nu};{\bf u})\cdots 
\hat\phi_{m_1-\frac12}^{+(3)}({\underline \nu};{\bf u})\times\\[3mm]
&\hat\phi_{m_2+\frac12}^{+(2)}({\underline \nu};{\bf u})
\hat\phi_{m_2+\frac32}^{+(2)}({\underline \nu};{\bf u})\cdots
\hat\phi_{m_1-\frac12}^{+(2)}({\underline \nu};{\bf u})| -m_1-\nu_1,
-m_1-\nu_2,-m_1-\nu_3\rangle\, .
\end{split}
\end{equation}

Let 
$
p=\max\{ m_1+\nu_i\}$,
then it is a straightforward calculation to check that
\begin{equation}
\label{6a}
\begin{split}
\hat G&({\underline \nu};{\bf u})|0\rangle=(-)^{\frac12 p(p+1)+m_1\nu_2+m_1p+p\nu_2+\nu_1\nu_2+\nu_1\nu_3+\nu_2\nu_3}
\hat\phi_{m_3+\frac12}^{+(3)}({\underline \nu};{\bf u})
\hat\phi_{m_3+\frac32}^{+(3)}({\underline \nu};{\bf u})\cdots\\[3mm]
&\cdots \hat\phi_{m_1-\frac12}^{+(3)}({\underline \nu};{\bf u})
\hat\phi_{m_2+\frac12}^{+(2)}({\underline \nu};{\bf u})
\hat\phi_{m_2+\frac32}^{+(2)}({\underline \nu};{\bf u})\cdots
\hat\phi_{m_1-\frac12}^{+(2)}({\underline \nu};{\bf u})\psi^{+(1)}_{m_1+\nu_1+\frac12}
\psi^{+(1)}_{m_1+\nu_1+\frac32}\cdots\\[3mm]
&\cdots\psi^{+(1)}_{p-\frac12}\psi^{+(2)}_{m_1+\nu_2+\frac12}
\cdots\psi^{+(2)}_{p-\frac12}
\psi^{+(3)}_{m_1+\nu_3+\frac12}
\cdots\psi^{+(3)}_{p-\frac12}
| -3p\rangle
\end{split}
\end{equation}
Note that in the expression (\ref{6a}) 
one of the terms (and possibly more)
\[
\psi^{+(i)}_{m_1+\nu_i+\frac12}
\psi^{+(i)}_{m_1+\nu_i+\frac32}\cdots\psi^{+(i)}_{p-\frac12}
\]
does not appear, viz. the term for which $p=m_1+\nu_i$. 
Also 
\[
\begin{split}
|-3p\rangle&=(Q_1^{-1}Q_2^{-1}Q_3^{-1})^p|0\rangle\\
&=v_{-3p-\frac12}\wedge v_{-3p-\frac32}\wedge v_{-3p-\frac52}\wedge\cdots\, .
\end{split}
\]
Hence, it is possible to express the tau function 
as a $3p\times 3p$-determinant.

\section{Conditions satisfied by the constructed tau function}

First we observe by using (\ref{5}) and (\ref{6}) that 
\begin{equation}
\label{7}
\begin{split}
\tau({\underline \nu};{\bf u})=0\quad &\mbox{if }\quad
\nu_1+\nu_2+\nu_3\ne -m_1-m_2-m_3=\mu_1+\mu_2+\mu_3\, ,\\ 
&\mbox{or one }\quad
\nu_i <-m_1\, ,\\
&\mbox{or one }\quad \nu_i>-m_3\, .
\end{split}
\end{equation}
Note that these equations (\ref{7}) define the complement of a convex polygon in the plane $x_1+x_2+x_3=\mu_1+\mu_2+\mu_3$. In other words, if one defines  $\mbox{supp }\tau$, the {\it support} of $\tau$, those $\underline\nu\in\mathbb{Z}^3$ for which  $\tau({\underline \nu};{\bf u})\ne 0$, then 
supp~$\tau$ is within a convex polygon (see also \cite{KL}).

{}From (\ref{5}) we deduce that
\[
\begin{split}
\sum_{a=1}^3\frac{\partial\hat \phi_\ell^{+(b)}({\underline \nu};{\bf u})}{\partial u_1^{(a)}}
=&
\sum_{a=1}^3\sum_{j=\ell+\nu_a}^{m_1+\nu_a-\frac12} f_a^{(b)}({\underline \nu})\psi_{j}^{+(a)}\frac{\partial S_{j-\ell-\nu_a}(u^{(a)})}{\partial u_1^{(a)}}\\
=&
\sum_{a=1}^3\sum_{j=\ell+\nu_a}^{m_1+\nu_a-\frac12} f_a^{(b)}({\underline \nu})\psi_{j}^{+(a)}S_{j-\ell-\nu_a-1}(u^{(a)})\\
=&\hat \phi_{\ell+1}^{+(b)}({\underline \nu};{\bf u})\, .
\end{split}
\]
Also observe  in (\ref{5}) that $\hat \phi_{m_1+\frac12}^{+(b)}({\underline \nu};{\bf u})=0$.
Hence from the explicit form of (\ref{6}), we conclude that
\begin{equation}
\label{ttau}
\hat I ( \tau({\underline \nu};{\bf u}))= \sum_{a=1}^3\frac{\partial\tau({\underline \nu};{\bf u})}{\partial u_1^{(a)}}=0\, .
\end{equation}

There exists a natural gradation on this Spin module with times $u_k^{(i)}$, which we denote by deg.
This gradation is the sum of two separate gradations, a ${\bf u}$-gradation $\mbox{deg}_{\bf u}$, which is given by 
$\mbox{deg}_{\bf u}(u_i^{(a)})=i$  and a fermionic gradation $\mbox{deg}_f$, which is defined by $\mbox{deg}_f(\psi_{\mathfrak k}^{\pm (a)})=-{\mathfrak k}$. So 
\[
\mbox{deg}=\mbox{deg}_{\bf u}+\mbox{deg}_f\, .
\]
Now,\[
\mbox{deg}_{\bf u} (S_j(u^{(a)}))=j
\]
 and the ${\bf u}$-gradation is related to
the Euler vector field $\hat E$, viz. for a homogeneous function $f$ one has 
\[
\mbox{deg}_{\bf u} (f({\bf u}))f({\bf u})=\sum_{a=1}^3\sum_{i=1}^\infty  iu_i^{(a)}\frac{\partial f({\bf u})}{\partial u_i^{(a)}}\, .
\]
Since, $\mbox{deg}\, |m_1,m_2,m_3\rangle=\frac12\left(m_1^2+m_2^2+m_3^2\right)$ and $\mbox{deg}\, \psi_k^{\pm (a)}=-k$,
it is straightforward to show that
 $\mbox{deg}\, G|0\rangle=\frac12\left( m_1^2+m_2^2+m_3^2\right)$, (see (\ref{1})). Thus,
we see from (\ref{1a}) that 
\[
\mbox{deg}\, g(\nu_1,\nu_2,\nu_3)=\mbox{deg}\, G|0\rangle -
\mbox{deg}\, |\nu_1,\nu_2,\nu_3\rangle=
\frac12\left(m_1^2+m_2^2+m_3^2-\nu_1^2-\nu_2^2-\nu_3^2\right)\, ,
\]
which by (\ref{eq:Roms}) is equal to $R^2$.
Hence, using (\ref{1b}), we find that
\begin{equation}
\label{dnu}
\mbox{deg}_{\bf u}(\tau({\underline\nu};{\bf u}))= R^2\, .
\end{equation}
Hence we can conclude that 
\begin{equation}
\label{tttau}
\sum_{a=1}^3\sum_{i=1}^\infty  iu_i^{(a)}\frac{\partial \tau({\underline\nu};{\bf u})}{\partial u_i^{(a)}}
=R^2 \tau({\underline\nu};{\bf u})\, .
\end{equation}
If one then puts 
all $u_i^{(a)}$, for all $i>1$,
in 
$\tau({\underline\nu};{\bf u})$ equal
to zero  and writes $\tau({\underline\nu};{u})$ for this tau function, equation (\ref{tttau})
reduces to
\begin{equation}
\label{ttttau}
\hat E(\tau({\underline\nu};{ u}))=
\sum_{a=1}^3u_1^{(a)}\frac{\partial \tau({\underline\nu};{u})}{\partial u_1^{(a)}}
=R^2\, \tau({\underline\nu};{ u})\, .
\end{equation}

\section{The tau function as a determinant}
Recall from (\ref{relabelv}) that $v_{j}^{(a)}=v_{3j-2+a}$.
In analogy with 
(\ref{5}) we define
\begin{equation}
\label{5v}
\begin{split}
v_\ell^{(b)}({\underline \nu};{\bf u})=&
\sum_{a=1}^3\sum_{j=\ell+\nu_a}^{m_1+\nu_a-\frac12} f_a^{(b)}({\underline \nu})S_{j-\ell-\nu_a}(u^{(a)})v_{j}^{(a)}\\[3mm]
=&\sum_{a=1}^3\sum_{j=\ell+\nu_a}^{m_1+\nu_a-\frac12} f_a^{(b)}({\underline \nu})S_{j-\ell-\nu_a}(u^{(a)})v_{3j-2+a}
\, .
\end{split}
\end{equation}
Hence, from (\ref{6a}) we obtain that
\begin{equation}
\label{6va}
\begin{split}
\hat G&({\underline \nu};{\bf u})|0\rangle=(-)^{\frac12 p(p+1)+m_1\nu_2+m_1p+p\nu_2+\nu_1\nu_2+\nu_1\nu_3+\nu_2\nu_3}
v_{m_3+\frac12}^{+(3)}({\underline \nu};{\bf u})\wedge v_{m_3+\frac32}^{+(3)}({\underline \nu};{\bf u})\wedge \cdots\\[3mm]
&\cdots \wedge v_{m_1-\frac12}^{+(3)}({\underline \nu};{\bf u})
\wedge v_{m_2+\frac12}^{+(2)}({\underline \nu};{\bf u})
\wedge v_{m_2+\frac32}^{+(2)}({\underline \nu};{\bf u})\cdots
\wedge v_{m_1-\frac12}^{+(2)}({\underline \nu};{\bf u})\wedge v^{(1)}_{m_1+\nu_1+\frac12}
\wedge\cdots\\[3mm]
&\cdots\wedge v^{(1)}_{p-\frac12}\wedge v^{(2)}_{m_1+\nu_2+\frac12}\wedge 
\cdots\wedge v^{(2)}_{p-\frac12}
\wedge v^{(3)}_{m_1+\nu_3+\frac12}
\cdots\wedge v^{(3)}_{p-\frac12}\wedge
v_{-3p-\frac12}\wedge
v_{-3p-\frac32}\wedge
\cdots
\end{split}
\end{equation}

Using a formula from \cite{Kac-Raina}, viz. (4.48), we find that
\[
\tau({\underline \nu};{\bf u})=(-)^{\frac12 p(p+1)+m_1\nu_2+m_1p+p\nu_2+\nu_1\nu_2+\nu_1\nu_3+\nu_2\nu_3}\det(A)\, ,
\]
where $A$ is the following $3p\times 3p$ matrix:
\begin{equation}
\begin{split}
\label{A}
A&=\sum_{i=1}^{m_1-m_3} \sum_{k=1}^{3}
f_k^{(3)}(\underline\nu)\sum_{j=0}^{m_1+\nu_k-1}
S_{j-m_3-\nu_k-i+1}(u^{(k)})E_{-3j+k-\frac72,-i+\frac12}\\
&+\sum_{i=1}^{m_1-m_2} \sum_{k=1}^{3} 
f_k^{(2)}(\underline\nu)\sum_{j=0}^{m_1+\nu_k-1}
S_{j-m_2-\nu_k-i+1}(u^{(k)})E_{-3j+k-\frac72,m_3-m_1-i+\frac12}
\\
&+\sum_{i=1}^{p-m_1-\nu_1}
E_{-3(m_1+\nu_1+i)+\frac12,m_3+m_2-2m_1+\frac12-i}\\
&+\sum_{i=1}^{p-m_1-\nu_2}
E_{-3(m_1+\nu_2+i)+\frac32,m_3+m_2-m_1-p+\nu_1+\frac12-i}\\
&+\sum_{i=1}^{p-m_1-\nu_3}
E_{-3(m_1+\nu_3+i)+\frac52,m_3+m_2-2p+\nu_1+\nu_2+\frac12-i}\, .
\end{split}
\end{equation}
Now replacing the indices such that  $E_{-i,-j}$ becomes $E_{i+\frac12, j+\frac12}$ and using some elementary matrix operations, see appendix \ref{A1} for the explicit calculations, we obtain that 
\[
\tau({\underline \nu};{\bf u})=(-)^{m_1\nu_2+\nu_1\nu_2+\nu_1\nu_3+\nu_2\nu_3}\det(E)\, ,
\]
where
\begin{equation}
\label{E}
\begin{split}
E=&w_1^{(3)}\sum_{j=1}^{m_1+\nu_1}
E_{2m_1-m_2-m_3-j+1,m_1-m_3-j+1}+\\[3mm]
&\qquad \sum_{i=1}^{m_1-m_3}\left(w_2^{(3)}\sum_{j=1}^{m_1+\nu_2}
S_{j-m_3-\nu_2-i}(u^{(2)}-u^{(1)})E_{m_1+\nu_3+j,i}+
\right.\\[3mm]
&\qquad \left. 
w_3^{(3)}\sum_{j=1}^{m_1+\nu_3}
S_{j-m_3-\nu_3-i}(u^{(3)}-u^{(1)})E_{j,i}
\right)+\\[3mm]
&w_1^{(2)}\sum_{j=1}^{\min\{ m_1+\nu_1, m_1-m_2\} }
E_{2m_1-m_2-m_3-j+1,2m_1-m_2-m_3-j+1}+
\\[3mm]
&\qquad \sum_{i=1}^{m_1-m_2}\left(w_2^{(2)}\sum_{j=1}^{m_1+\nu_2}
S_{j-m_2-\nu_2-i}(u^{(2)}-u^{(1)})E_{m_1+\nu_3+j,-m_3+m_1+i}+\right.\\[3mm]
&\qquad \left. w_3^{(2)}\sum_{j=1}^{m_1+\nu_3}
S_{j-m_2-\nu_3-i}(u^{(3)}-u^{(1)})E_{j,-m_3+m_1+i}
\right)
\, .
\end{split}
\end{equation}
To make the connection with the Painlev\'e VI tau-function, we put all higher times $u_n^{(a)}$ for $n>0$ to zero.
Then $S_k(u^{(a)}-u^{(b)})$ becomes $\frac{(u_a-u_b)^k}{k!}$. Using (\ref{th}) we can express this tau function in 
$t$ and $h$, viz. we substitute
\[
u_2-u_1=h, \qquad u_3-u_1=\frac{h}{t}\,.
\]
Since $\tau(\underline \nu;{\bf u})$ is homogeneous of $\bf u$-degree $R^2$ (see (\ref{dnu}),
we find that
\[
\tau(\underline \nu;t,h)=h^{R^2}\tau_0(\underline \nu;t)
\]
and
\[
\tau_0(\underline \nu;t)=\frac{(-)^{m_1\nu_2+\nu_1\nu_2+\nu_1\nu_3
+\nu_2\nu_3}}{t^{R^2}}\det(T_{\underline \nu})\, ,
\]
where 
\begin{equation}
\label{TTT}
\begin{split}
T_{\underline \nu}=&w_1^{(3)}\sum_{j=1}^{m_1+\nu_1}
E_{2m_1-m_2-m_3-j+1,m_1-m_3-j+1}+\\[3mm]
&\qquad \sum_{i=1}^{m_1-m_3}\left(w_2^{(3)}\sum_{j=1}^{m_1+\nu_2}
t^{(j-m_3-\nu_2-i)}E_{m_1+\nu_3+j,i}+
w_3^{(3)}\sum_{j=1}^{m_1+\nu_3}
1^{(j-m_3-\nu_3-i)}E_{j,i}
\right)+\\[3mm]
&w_1^{(2)}\sum_{j=1}^{\min\{ m_1+\nu_1, m_1-m_2\} }
E_{2m_1-m_2-m_3-j+1,2m_1-m_2-m_3-j+1}+
\\[3mm]
&\qquad \sum_{i=1}^{m_1-m_2}\left(w_2^{(2)}\sum_{j=1}^{m_1+\nu_2}
t^{(j-m_2-\nu_2-i)}E_{m_1+\nu_3+j,-m_3+m_1+i}+
\right.\\[3mm]
&\qquad \left. 
w_3^{(2)}\sum_{j=1}^{m_1+\nu_3}
1^{ ( j-m_2-\nu_3-i )}E_{j,-m_3+m_1+i}
\right)
\, .
\end{split}
\end{equation}
Here 
$t^{(n)}$ stands for the divided power
\[
t^{(n)}=\frac{t^{n}}{n!}\, .
\]
Note that we use the convention that $n!=\infty$ for $n<0$, such that $1^{(n)}=\frac1{n!}=t^{(n)}= \frac{t^n}{n!}=0$.
\section{An example}
Now assume that $\mu_1=-4$, $\mu_2=-2$, $\mu_3=0$, $\nu_1=-3$, 
$\nu_2=-2$ and $\nu_3=-1$. Then $R^2=3$.
{}From \rf{posvis} we find 
\begin{equation}
v_1=2,\quad v_2=0,\quad v_3=-2\quad\mbox{ and }v_4=-1\, .
\label{eq:exmpl1}
\end{equation} 
and equation \rf{param1} with $(i,j,k)=(3,2,1)$ 
yields for the PVI parameters :
\[
\alpha=\frac12,\quad \beta=-2,\quad \gamma=2,\quad\mbox{ and }
\delta=-\frac32\, .
\]
Then
\[
\tau=-(h/t)^3\det
\left(
\begin{array}{cccc|cc}
 w_3^{(3)} & w_3^{(3)}& 0 & 0 & 0 & 0 \\[3mm]
 \frac{ w_3^{(3)}}{2 } &w_3^{(3)} & w_3^{(3)} & 0 &w_3^{(2)} & 0 \\[3mm]
 \frac{w_3^{(3)}}{6 } & \frac{w_3^{(3)}}{2 } & w_3^{(3)} & w_3^{(3)} & w_3^{(2)} & w_3^{(2)} \\[2mm]
\hline \break
 \frac{t^2 w_2^{(3)}}{2} & t w_2^{(3)} & w_2^{(3)} & 0 & w_2^{(2)} & 0 \\[3mm]
 \frac{t^3 w_2^{(3)}}{6} & \frac{t^2 w_2^{(3)}}{2} & t w_2^{(3)} & w_2^{(3)} & tw_2^{(2)} & w_2^{(2)} \\[2mm]
\hline \break 
 0 & 0 & 0 & w_1^{(3)} & 0 & w_1^{(2)}
\end{array}\right)\, .
\]
Thus
\[
\tau_0=\frac{D}{6 t^3}
\left( D_1 + 
    3D_2 t^2    - 
    2D_2t^3 \right)\, ,
\]
where
\[ 
D=
w_3^{(3)} ( w_2^{(2)} w_3^{(3)}-w_2^{(3)} w_3^{(2)} ) 
\]
and 
\[
D_1=w_3^{(3)}(w_1^{(2)} w_2^{(3)} -w_1^{(3)} w_2^{(2)})
\qquad
D_2=w_2^{(3)} (w_1^{(3)} w_3^{(2)} - w_1^{(2)} w_3^{(3)})
\, .
\]
Then
\[
\sigma=2 \frac{D_1 -2D_1t-2D_2t^3+D_2t^4 }{ D_1+3D_2t^2 -
   2 D_2t^3  }
\]
and according to \rf{okay} this gives a Painlev\'e IV solution 
(\ref{y}), presented in the introduction.

{}From \rf{negvis} and \rf{param2} we get
\[
v_1=-2,\quad v_2=0,\quad v_3=2\quad\mbox{ and }v_4=-1\, .
\]
together with
\[
\alpha=\frac92,\quad \beta=-2,\quad \gamma=2,
\quad\mbox{ and }\delta=-\frac32\, .
\]
This time the solution of the Painlev\'e VI equation \rf{P6} takes the
form
\[
\begin{split}
y &= \,\frac13\,\left( -{\it D_1}+2\,{\it D_1}\,t+2\,{\it D_2}\,{t}^{3}-{\it
D_2}\,{t}^{4} \right)^{-1}\\
&\left( {{\it D_1}}^{2}-
6\,{\it D_1}\,{\it D_2}\,{t}^{2}+4\,{\it D_1}\,{\it D_2}\,{t}^{3}
+{{\it D_2}}^{2}{t}^{4}+4\,{\it D_2}\,t{\it D_1} \right)^{-1}\\
&\left( -{{\it D_2}}^{3}{t}^{9}+3\,{{\it D_2}}^{3}{t}^{8}-6\,{\it D_1}\,
{{\it D_2}}^{2}{t}^{7}
\right.\\
&\left.+42\,{\it D_1}\,{{\it D_2}}^{2}{t}^{6}-57\,{\it D_1}\,{{\it D_2}}^{2}{t}^{5}+27\,{{\it D_1}}^{2}{\it D_2}\,{t}^{5}
+27\,{\it D_1}\,{{\it D_2}}^{2}{t}^{4} \right.
\\
&\left.-57\,{{\it D_1}}^{2}{\it D_2}\,{t}^{4}+
42\,{{\it D_1}}^{2}{\it D_2}\,{t}^{3}-
6\,{{\it D_1}}^{2}{\it D_2}\,{t}^{2}+
3\,t{{\it D_1}}^{3}-{{\it D_1}}^{3} \right) \\
\end{split}
\]
as dictated by relation \rf{okay}.

Interchanging $v_2$ with $v_3$ in equation \rf{exmpl1},
which can be done by choosing $\mu_2=0$ and $\mu_3=-2$ (this does not change the tau-function), yields
\[
\alpha=\frac12,\quad \beta=0,\quad \gamma=8,
\quad\mbox{ and }\delta=\frac12\, .
\]
according to equation \rf{param1} with $(i,j,k)=(2,1,3)$ .
The corresponding solution of the Painlev\'e VI equation \rf{P6}
reads:
\[
\begin{split}
y&= \,\left( {\it D_1}+{\it D_2}\,{t}^{3}-3\,{\it D_2}\,{t}^{2}+
3\,{\it D_2}\,t \right)^{-1} \left( {{\it D_1}}^{2}-6\,{\it D_1}\,{\it
D_2}\,{t}^{2}\right.\\
&\left.+4\,{\it D_1}\,{\it D_2}\,{t}^{3}+{{\it D_2}}^{2}{t}^{4}+
4\,{\it D_2}\,{\it D_1}\,t \right)^{-1} \left({\it D_2}\,t \right)\\
& \left( -15\,{t}^{2}{{\it D_1}}^{2}+7\,t{{\it D_1}}^{2}+
9\,{t}^{3}{{\it D_1}}^{2}-{{\it D_1}}^{2}+
6\,{\it D_1}\,{\it D_2}\,{t}^{2}-26\,{\it D_1}\,{\it D_2}\,{t}^{3}
\right.\\
&\left.-6\,{\it D_1}\,{\it D_2}\,{t}^{5}
+26\,{\it D_1}\,{\it D_2}\,{t}^{4}
-9\,{{\it D_2}}^{2}{t}^{4}-7\,{{\it D_2}}^{2}{t}^{6}
+
15\,{{\it D_2}}^{2}{t}^{5}+{{\it D_2}}^{2}{t}^{7} \right) 
\end{split}
\]

\section{Wave matrix}
We want to calculate the wave matrix corresponding to $\underline\nu$. For this we let the fermionic field 
$\psi^{+(j)}(z)$ act on $G|0\rangle$. Then
\[
\begin{split}
\sigma\left(\psi^{+(j)}(z)G|0\rangle\right)=&
\sigma\left(\psi^{+(j)}(z)
\sum_{\underline \nu\in {\rm supp }\  \tau}g(\underline\nu)|
\underline\nu\rangle
\right)
\\[3mm]
=\sum_{\underline \nu\in {\rm supp }\ \tau}(-)^{\nu_1+\cdots +\nu_{j-1}}z^{\nu_j}&
e^{\sum_k u_k^{(j)}z^j}e^{-\sum_{k=1}^\infty\frac{\partial }{\partial u_k^{(j)}}\frac{z^{-k}}{k}}\tau(\underline \nu; {\bf u})
|\nu_1+\delta_{j1},\nu_2+\delta_{j2},
\nu_3+\delta_{j3}\rangle
\end{split}
\]
Now let $\Phi_{ij}(\underline\nu;{\bf u})$ be the  coefficient of
$|\nu_1+\delta_{i1},\nu_2+\delta_{i2},
\nu_3+\delta_{i3}\rangle$ of this expression, i.e.,  this term corresponds to 
\[
\langle \nu_1+\delta_{i1},\nu_2+\delta_{i2},
\nu_3+\delta_{i3}|
\psi^{+(j)}(z)G|0\rangle\, ,
\]
then
\[
\begin{split}
\Phi_{ij}(\underline\nu;{\bf u})&={sign}(i-j)(-)^{\nu_1+\cdots +\nu_{j-1}}z^{\nu_j}z^{\delta_{ij}-1}z^{\nu_j}e^{\sum_k u_k^{(j)}z^j}\times\\
&\qquad \times e^{-\sum_{k=1}^\infty\frac{\partial }{\partial u_k^{(j)}}\frac{z^{-k}}{k}}\tau(\nu_1+\delta_{i1}-\delta_{j1},\nu_2+\delta_{i2}-\delta_{j2},\nu_3+\delta_{i3}-\delta_{j3};{\bf u})
\end{split}
\]

For every $\underline\nu\in\mbox{supp }\tau$ we define a wave matrix $\Psi(\underline\nu;{\bf u})$ as follows
\begin{equation}
\label{8}
\begin{split}
\Psi(\underline\nu;{\bf u},z)&=\Theta(\underline\nu;{\bf u},z)\exp \left(\sum_{j=1}^3\sum_{n=1}^\infty z^nE_{jj}u_n^{(j)}\right)z^{\underline{  \nu}}\, ,\\[3mm]
\Psi(\underline\nu;{\bf u},z)&=\left(\frac{\Phi_{ij}(\underline\nu;{\bf u},z)
}{\tau(\underline\nu;{\bf u})}
\right)_{1\le ij\le 3}\, ,\\[3mm]
\Theta(\underline\nu;{\bf u},z)&=\left(\Theta(\underline\nu;{\bf u},z)_{ij}\right)_{1\le ij\le 3} \quad\mbox{with}\\[3mm]
\tau(\underline\nu
;{\bf u})\Theta_{ij}(\underline\nu;{\bf u},z)&=\\
\mbox{sign}(i-j)z^{\delta_{ij}-1}&
\exp\left(-\sum_{k=1}^\infty\frac{\partial}{\partial u_k^{(j)}}\frac{z^{-k}}{k}
\right)\tau(\nu_1+\delta_{i1}-\delta_{j1},\nu_2+\delta_{i2}-\delta_{j2},\nu_3+\delta_{i3}-\delta_{j3};{\bf u})\, .
\end{split}
\end{equation}
The coefficient of $z^{-1}$ of $\Theta(\underline\nu;{\bf u},z)$ is the matrix 
$\theta^{(-1)}(\underline\nu;{\bf u})$, see (\ref{eq:tgrad-exp}).
The off-diagonal elements are the rotation coefficients $\beta_{ij}(\underline\nu;{\bf u})$ (see 
(\ref{eq:them1})).
Now
\begin{equation}
\label{bbbeta}
\beta_{ij}(\underline\nu;{\bf u})=
\mbox{sign}(i-j)
\frac{
\tau(\nu_1+\delta_{i1}-\delta_{j1},\nu_2+\delta_{i2}-\delta_{j2},\nu_3+\delta_{i3}-\delta_{j3};{\bf u})}{\tau(\nu_1,\nu_2,\nu_3;{\bf u})}
\end{equation}
Using (\ref{dnu}) it is easy to  deduce that
 \[
 \mbox{deg}_{\bf u}(\beta_{ij}({\underline 
\nu};{\bf u}))=-1-\nu_i+\nu_j\, .
 \]
If we then put in (\ref{bbbeta}) all higher times $u_n^{(a)}=0$ for $n>0$ and make the substitution 
(\ref{th}), then for $k\ne i,j$ we find that
\begin{equation}
\label{bbeta}
\begin{split}
&\beta_{ij}(\underline\nu;t,h)=\\
&-\mbox{sign}(i-j)
(-)^{m_1(\delta_{i2}+\delta_{j2})
+\nu_1
+\nu_2
+\nu_3+\nu_k}
\left(\frac{h}{t}\right)^{\nu_j-\nu_i-1}
\frac{\det (T_{(
\nu_1+\delta_{i1}-\delta_{j1},\nu_2+\delta_{i2}-\delta_{j2},\nu_3+\delta_{i3}-\delta_{j3})})}{
\det(T_{(\nu_1,\nu_2,\nu_3)})}
\end{split}
\end{equation}

Looking at (\ref{1}) it is clear that for $j=1,2,3$:
\[
\phi^{+(j)}_k G|0\rangle=\begin{cases}0&\qquad\mbox{for }k>m_j\, ,\\
\ne0&\qquad\mbox{for }k<m_j\, ,
\end{cases}
\]
or in other words
\[
z^{m_j}\sum_{\ell=1}^3 w_\ell^{(j)}\psi^{+(\ell)}(z)G|0\rangle
\]
is a non-negative power series in $z$.
It is easy to see that in fact for $i=1,2,3$:
\[
\langle \nu_1+\delta_{i1},\nu_2+\delta_{i2},\nu_3+\delta_{i3}|z^{m_j}\sum_{\ell=1}^3 w_\ell^{(j)}\psi^{+(\ell)}(z)G|0\rangle
\]
is a non-negative power series in $z$, with nonzero constant term.
Using the boson-fermion correspondence this term corresponds to
\[
z^{m_j}\sum_{\ell=1}^3 w_\ell^{(j)}\Phi_{i\ell}(\underline\nu;{\bf u})\, .
\]
Thus we have shown that
\[
z^{m_j}\Psi(\underline\nu;{\bf u},z)\begin{pmatrix}
w_1^{(j)}\\ w_2^{(j)}\\ w_3^{(j)}
\end{pmatrix}
\]
contains no negative powers of $z$.
Stated in another way this means that
\[
\Pi(\underline\nu;{\bf u},z):=\Psi(\underline\nu;{\bf u},z)g_0 z^{-\underline\mu}
\]
has no negative powers of $z$ and the constant coefficient is an invertible matrix $M(\underline\nu;{\bf u})$.
\section{Conclusion/Outlook}
In summary we have shown how the general Painlev\'e VI equation
and its class of solutions are obtained from a self-similarity
reduction of  a $3$-component KP hierarchy.
We have shown how reduction is imposed through restricting
group elements $g$ in the Riemann-Hilbert
factorization to be of special form determined by five independent
scaling parameters. These parameters specify the scaling laws of
the underlying dressing matrices and values of Painlev\'e VI
coefficients through simple relations.

The particular advantage of our construction is that it
allows for an explicit construction of the tau function
solutions of the sigma-form of the  Painlev\'e VI
equation in terms of elementary Schur functions
using Grassmannian techniques.

The approach outlined in this paper offers powerful techniques
to study symmetries of the Painlev\'e VI equation
which originate in the formalism of B\"acklund transformations
within the setup of  a $3$-component KP hierarchy.
The work is in progress to develop detailed and exhaustive 
results within this conceptual framework.

\appendix
\section{Appendix: Calculation of the tau function}
\label{A1}
Consider the matrix $A$ given in (\ref{A}), replace the indices such that  $E_{-i,-j}$ becomes $E_{i+\frac12, j+\frac12}$ then $A$ has the form
\begin{equation}
\label{AA}
\begin{split}
A=&\sum_{i=1}^{m_1-m_3}\sum_{k=1}^3 f_k^{(3)}(\underline\nu)
\sum_{j=0}^{m_1+\nu_k-1} S_{j-m_3-\nu_k-i+1}(u^{(k)})E_{3j+4-k,i}
\\
&+\sum_{i=1}^{m_1-m_2} \sum_{k=1}^3  f_k^{(2)}(\underline\nu)
\sum_{j=0}^{m_1+\nu_k-1}
S_{j-m_2-\nu_k-i+1}(u^{(k)})E_{3j+4-k,-m_3+m_1+i}\\
&+\sum_{i=1}^{p-m_1-\nu_1}
E_{3(m_1+\nu_1+i),-m_3-m_2+2m_1+i}\\
&+\sum_{i=1}^{p-m_1-\nu_2}
E_{3(m_1+\nu_2+i)-1,-m_3-m_2+m_1+p-\nu_1+i}\\
&+\sum_{i=1}^{p-m_1-\nu_3}
E_{3(m_1+\nu_3+i)-2,-m_3-m_2+2p-\nu_1-\nu_2+i}\, .
\end{split}
\end{equation}
Next permute the rows of this matrix, then we obtain that 
\[
\tau({\underline \nu};{\bf u})=(-)^{p+m_1\nu_2+m_1p+p\nu_2+\nu_1\nu_2+\nu_1\nu_3+\nu_2\nu_3}\det(B)\, ,
\]
where 
\begin{equation}
\label{B}
\begin{split}
B&=\sum_{i=1}^{m_1-m_3}\sum_{k=1}^3 
f_k^{(3)}(\underline\nu)\sum_{j=0}^{m_1+\nu_k-1}
S_{j-m_3-\nu_k-i+1}(u^{(k)})E_{(3-k)p+j+1,i}\\
&+\sum_{i=1}^{m_1-m_2}\sum_{k=1}^3 
f_k^{(2)}(\underline\nu)\sum_{j=0}^{m_1+\nu_k-1}
S_{j-m_2-\nu_k-i+1}(u^{(k)})E_{(3-k)p+j+1,-m_3+m_1+i}
\\
&+\sum_{i=1}^{p-m_1-\nu_1}
E_{2p+m_1+\nu_1+i,-m_3-m_2+2m_1+i}\\
&+\sum_{i=1}^{p-m_1-\nu_2}
E_{p+m_1+\nu_2+i,-m_3-m_2+m_1+p-\nu_1+i}\\
&+\sum_{i=1}^{p-m_1-\nu_3}
E_{m_1+\nu_3+i,-m_3-m_2+2p-\nu_1-\nu_2+i}\, .
\end{split}
\end{equation}
Now developing the determinant to the last columns we obtain that
\[
\tau({\underline \nu};{\bf u})=(-)^{m_1\nu_2+\nu_1\nu_2+\nu_1\nu_3+\nu_2\nu_3}\det(C)\, ,
\]
where
\begin{equation}
\label{C}
\begin{split}
C&=\sum_{i=1}^{m_1-m_3}\sum_{k=1}^3
f_k^{(3)}(\underline\nu)\sum_{j=1}^{m_1+\nu_k}
S_{j-m_3-\nu_k-i}(u^{(k)})E_{(3-k)m_1+\sum_{p>k}\nu_p+j,i}
\\
&+\sum_{i=1}^{m_1-m_2}
\sum_{k=1}^3 f_k^{(2)}(\underline\nu)\sum_{j=1}^{m_1+\nu_k}
S_{j-m_2-\nu_k-i}(u^{(k)})E_{(3-k)m_1+\sum_{p>k}\nu_p+j,-m_3+m_1+i}
\, .
\end{split}
\end{equation}
Now note that $\det(C)=\det(D)$, 
\begin{equation}
\label{D}
\begin{split}
D&=\sum_{i=1}^{m_1-m_3}\sum_{k=1}^3
w_k^{(3)}\sum_{j=1}^{m_1+\nu_k}
S_{j-m_3-\nu_k-i}(u^{(k)})E_{(3-k)m_1+\sum_{p>k}\nu_p+j,i}
\\
&+\sum_{i=1}^{m_1-m_2}\sum_{k=1}^3
w_k^{(2)}\sum_{j=1}^{m_1+\nu_k}
S_{j-m_2-\nu_k-i}(u^{(k)})E_{(3-k)m_1+\sum_{p>k}\nu_p+j,-m_3+m_1+i}
\, .
\end{split}
\end{equation}
Next multiply $D$ from the right with the matrix (with determinant 1):
\[
\sum_{i,j=1}^{m_1-m_3} S_{i-j}(-u^{(1)})E_{ij}+ 
\sum_{i,j=1}^{m_1-m_2} S_{i-j}(-u^{(1)})E_{m_1-m_3+i,m_1-m_3+j}
\]
Thus $\det(D)=\det(E)$, for $E$ as in (\ref{E}). Here we have used that $3m_1+\nu_1+\nu_2+\nu_3=2m_1-m_2-m_3$.
\vskip .4cm \noindent
{\bf Acknowledgments} \\
Work of JvdL is partially supported by the European
 Union through the FP6
Marie Curie RTN {\em ENIGMA} (Contract number MRTN-CT-2004-5652)
and the European Science Foundation Program MISGAM.

\noindent
Work of HA is partially supported by grant NSF PHY-0651694.


\bigskip

\begin{thebibliography}{99}

\bibitem{Aratyn:2001cj}
  H.~Aratyn and J.~van de Leur,
  Commun.\ Math.\ Phys.\  { 239}, 155 (2003)
  [arXiv:hep-th/0104092].

\bibitem{Aratyn:2001jv}
  H.~Aratyn and J.~van de Leur,
  Theor.\ Math.\ Phys.\  { 134}, 14 (2003)
  [Teor.\ Mat.\ Fiz.\  { 134}, 18 (2003)]
  [arXiv:hep-th/0111243].

\bibitem{Aratyn:2004nz}
  H.~Aratyn and J.~van de Leur,
Annales de l'Institut Fourier { 55}, 1871-1903 (2005), 
[arXiv:nlin.SI/0406038]

\bibitem{Aratyn:2006vc}
  H.~Aratyn and J.~van de Leur,
SIGMA { 3}, 020 (2007)   [arXiv:nlin.si/0605027]


\bibitem{Boalch}
P. Boalch, in {\sl Theories asymptotiques et equations de Painlev\'e - Angers},
Eric Delabaere - Michèle Loday-Richaud (Ed.)
SMF, Seminaires et congres, vol 14, (2006) 1--20 [arXiv:math.AG/0503043]

\bibitem{conte}
R. Conte, A. M. Grundland and M. Musette,
J. Phys. A: Math. Gen. 39,  12115-12127 (2006),   
 [arXiv:nlin.SI/0604011]

\bibitem{cosgrove}
C. M. Cosgrove,   J. Phys. A: Math. Gen. 39 (2006) 11955-­11971;
Stud. Appl. Math. 104 171-­228 (2000) 

\bibitem{DJKM1}
E. Date, M. Jimbo, M. Kashiwara and T. Miwa, 
J. Phys. Soc. Japan 50,  3806-3812 (1981).

\bibitem{Dubrovin:1998fe}
B.~Dubrovin,
Painlev\'e transcendents and two-dimensional topological field theory,
in ``{\sl The Painlev\'e property: 100 years later}'', 287-412, CRM Ser. Math.Phys,
Springer, New York 1999, [arXiv:math.ag/9803107].

\bibitem{Flaschka:1980wj}
  H.~Flaschka and A.~C.~Newell,
  Commun.\ Math.\ Phys.\  { 76}, 65-116 (1980).

\bibitem{Harnad}
J.~Harnad, 
Comm. Math. Phys. {166}, 337--365 (1994).


\bibitem{JM} 
M. Jimbo and T. Miwa, 
Physica 2D, 407-448 (1981).

\bibitem{KL}V. G. Kac and J. W. van de Leur,
Jour. Math. Phys. 44,  3245--3293 (2003).


\bibitem{Kac-Raina}
V.G. Kac and A.K. Raina, {\sl Bombay lectures on highest weight 
representations},  Adv. Ser. Math. Phys. 2, Singapore: World Scientific, 1987



\bibitem{kakei}
S. Kakei and T. Kikuchi,
Lett. Math. Phys. 79, 221--234 (2007),   [arXiv:nlin.SI/0508021]
 
\bibitem{joshi}
N. Joshi, A. V. Kitaev and P. A. Treharne. 
J. Math. Phys. 48, 103512 (2007) [arXiv:math.CA/0706.1750]



\bibitem{TV} F. ten Kroode and J. van de Leur, 
Comm. Math. Phys. 137,  67--107 (1991).

\bibitem{vdLeur-Martini}
J.W. van de Leur and R. Martini, 
Comm. Math. Phys. 205,  587--616 (1999)

\bibitem{M2002}
M.~Mazzocco,
Painlev\'{e} sixth equation as isomonodromic
deformations equation of an irregular system,
in ``{\sl The Kowalevski property}'', CRM Proc. Lecture Notes {\bf32}
(2002), Providence, RI 219--238.


\bibitem{mazzoccomo}
M.~Mazzocco and M.Y.~Mo,
Nonlinearity { 20}, 2845--2882 (2007), [arXiv:nlin.SI/0610066]

\bibitem{Okamoto}
K. Okamoto, 
Annali di Mathematica pura ed applicata { 146},  337-381 (1987) 

\bibitem{takasaki}
K. Takasaki,
SIGMA { 3}, 042 (2007),  [arXiv:nlin.SI/0610073]
\end{thebibliography}
\end{document}